\documentclass[runningheads]{llncs}

\usepackage[T1]{fontenc}

\usepackage{marvosym}
\usepackage{tikz}
\usetikzlibrary{shapes.geometric}
\usetikzlibrary{matrix}
\usepackage{subcaption}
\usepackage{caption}
\usepackage{fontawesome}

\usepackage{amsmath}
\usepackage{amsfonts}
\usepackage{amssymb}
\usepackage{environ}
\usepackage{mathtools}
\usepackage{xspace}
\usepackage{xparse}
\usepackage{stackrel}
\usepackage{tikz}
\usetikzlibrary{decorations.pathreplacing, positioning, arrows.meta}
\usepackage[Tol, keep-defaults]{colorblind}
\usepackage{cleveref}

\newcommand{\cw}{\textsc{Cutwidth}\xspace}
\DeclareDocumentCommand{\cwi}{O{i}m}{\ensuremath{\mathbf{cw}_{#1}(#2)}}
\newcommand{\Pred}{\text{Pred}}

\newcommand{\ZZ}{\mathbb{Z}}
\newcommand{\bN}{\mathbb{N}}
\newcommand{\bZ}{\mathbb{Z}}
\newcommand{\cC}{\mathcal{C}}
\newcommand{\cT}{\mathcal{T}}
\newcommand{\ext}{\text{ext}}

\newcommand{\type}{T}

\DeclareRobustCommand{\symClique}{
  \tikz[baseline=-0.6ex]
    \fill[black] (0,0.8ex)--(-0.7ex,-0.6ex)--(0.7ex,-0.6ex)--cycle;}

\DeclareRobustCommand{\symCliqueVert}{
  \tikz[baseline=-0.6ex]\fill[black] (0,0) circle (0.6ex);}

\DeclareRobustCommand{\symTC}{
  \tikz[baseline=-0.6ex]{
    \fill[black] (0,0) circle (0.6ex);
    \draw[black, line width=0.25pt] (0,0) circle (0.9ex);
  }
}

\usepackage{colonequals}

\def\coloniff{\mathrel{\ratio\Longleftrightarrow}}

\DeclareMathOperator\maxcut{maxcut}

\DeclarePairedDelimiter\abs{\lvert}{\rvert}
\DeclarePairedDelimiter\norm{\lVert}{\rVert}
\DeclarePairedDelimiterX\setImpl[1]{\{}{\}}{#1}
\NewDocumentCommand\set{sO{}m>{\TrimSpaces}o}{
  \IfValueTF{#4}{
    \setImpl[#2]{#3 : \IfBooleanTF{#1}{\text{#4}}{#4}}
  }{
    \setImpl[#2]{#3}
  }
}

\DeclareDocumentCommand\complexityclassfont{}{\sffamily}
\DeclareDocumentCommand\DeclareComplexityClass{mm}{\DeclareDocumentCommand{#1}{}{\textup{\complexityclassfont #2}\xspace}}
\DeclareComplexityClass\FPT{FPT}
\DeclareComplexityClass\XP{XP}
\DeclareComplexityClass\NP{NP}
\DeclareComplexityClass\coNP{coNP}

\newcommand{\restate}[2]{
  \begingroup
  \expandafter\renewcommand\csname the#1\endcsname{\ref{#2}}
  \expandafter\expandafter\expandafter\expandafter\expandafter\expandafter\csname #1\endcsname
  \csname reptheorem@#2\endcsname
  \csname end#1\endcsname
  \endgroup
}

\NewEnviron{restateable}[2]{%
  \global\expandafter\xdef\csname reptheorem@#2\endcsname{
    \unexpanded\expandafter{\BODY}
  }
  \expandafter\expandafter\csname #1\endcsname\BODY\unskip\label{#2}\csname end#1\endcsname
}

\spnewtheorem{observation}{Observation}{\bfseries}{\itshape}
\crefname{observation}{Observation}{Observations}

\begin{document}
\author{Klara Jacobsen\inst{1}\orcidID{0009-0008-9910-1544} \and
Klaus Jansen\inst{1}\orcidID{0000-0001-8358-6796} \and
Lis Pirotton\inst{1}\orcidID{0009-0001-4984-3696} \and
Björn Schumacher\inst{1}\orcidID{0009-0002-2448-4794}}

\authorrunning{K. Jacobsen et al.}
\institute{Kiel University, Kiel, Germany \\\email{stu241490@mail.uni-kiel.de, \{kj,lpi,bsch\}@informatik.uni-kiel.de}}

\title{New Insights into the Parameterized Complexity of \textsc{Cutwidth}}
\maketitle

\begin{abstract}
    The \cw problem asks for a linear ordering of the vertices that minimizes the maximum number of edges crossing between two consecutive positions.
    The problem is proven to be \NP-hard by Gavril [CISS '77], which motivated research in parameterized complexity.
    A previously claimed \FPT result for \cw and \textsc{Imbalance} parameterized by twin cover by Gorzny and Buss [COCOON '19] was later invalidated. Misra and Mittal [Theor. Comput. Sci. '21] clarified this issue, proved that \textsc{Imbalance} is in \XP parameterized by twin cover and in \FPT parameterized by $\omega$-bounded twin cover, i.e., where the size of the cliques outside the twin cover are bounded by $\omega$. The analogous questions remained open for cutwidth.  
    In this work, we answer one of them by showing that \cw is \FPT parameterized by $\omega$-bounded twin cover. 
    Additionally, we present an \XP algorithm for \cw parameterized by vertex integrity and prove $W[1]$-hardness for the parameter treedepth.
\end{abstract}

\section{Introduction}
The \cw problem is a fundamental graph layout problem.
We are given a graph $G=(V,E)$ with $\abs{V}= n$ and we seek a linear ordering of the vertices, defined as a bijective function $\pi: V \to \{1,\dots,n\}$, that minimizes the maximum number of edges crossing a line between any two consecutive vertices.
More formally, let $\Pi_G$ be the set of all orderings of $G$.
The \emph{maximum-cut} of $G$ associated with a permutation $\pi \in \Pi_G$ is defined as
\begin{align*}
     \maxcut_G(\pi) \coloneqq& \max_{i \in [n-1]}|\{\{u,v\}\in E : \pi(u) \leq i < \pi(v)\}|.
\end{align*}
The \emph{cutwidth value} of a graph is the minimum possible maximum-cut over all its linear layouts, i.e., $\mathbf{cw}(G) \coloneqq \min_{\pi \in \Pi_G} \maxcut_G(\pi)$.
\begin{definition}[\cw]
    Given a graph $G$, the \cw problem asks for the optimal cutwidth value $\mathbf{cw}(G)$.
\end{definition}
 \cw has numerous applications ranging from theoretical bounds to VLSI and bioinformatics~\cite{blin08,Breuer77,Chung85,Djidjev02,Leiserson80}.

The problem is \NP-hard~\cite{Gavril11}, even for restricted graph classes like planar graphs of maximum degree 3~\cite{Monien88} or split graphs, and consequently also chordal graphs~\cite{Heggernes11}.
The currently best exact algorithms for \cw without any additional structure are proposed by Bodlaender, Fomin, Koster, Kratsch, and Thilikos~\cite{BodlaenderFKKT12}.

\begin{figure}[t]
  \centering
  \begin{tikzpicture}[
    on grid,
    param/.style={minimum height=.5cm, draw=black, font=\small, align=center, rounded corners, semithick},
    fpt/.style={param, fill=T-Q-M2},
    xp/.style={param, fill=T-Q-M4!60},
    open/.style={param},
    whard/.style={param, fill=T-Q-M6},
    nphard/.style={param, fill=T-Q-M9!60},
    >={Stealth[round]}]
    \node[fpt] (vc) {vertex cover\\number \cite{FellowsLMRS08}};
    \node[fpt, left=4cm of vc] (ecc) {edge clique\\cover number \cite{gorzny22phd}};
    \node[fpt, below left=1.5cm and 1.75cm of vc] (rtc) {restricted twin\\cover number \cite{gorzny22phd}};
    \node[fpt, font={\small\bfseries}, below right=1.5cm and 1.6cm of vc] (obtc) {\(\omega\)-bounded twin \\cover {[\Cref{thm:FPT}]}};
    \node[open, below=1.5cm of rtc] (tc) {twin cover\\number};
    \node[xp, below=3cm of ecc] (nd) {neighborhood\\diversity \cite{Bakken18}};
    \node[open, below=2.5cm of nd] (mw) {modular width};
    \node[xp, font={\small\bfseries}, below=1.5cm of obtc] (cvdo) {$\omega$-bounded cluster\\ vertex deletion set};
    \node[open, below left=4.5cm and 1cm of vc] (cvd) {cluster vertex\\deletion number};
    \node[whard, font={\small\bfseries}, below left=2cm and .5cm of cvd] (sd) {shrub-depth};
    \node[nphard, below =2cm of sd] (cw) {cliquewidth};

    \node[whard, font={\small\bfseries}, above right=.5cm and 3.4cm of sd, align=center] (td) {treedepth\\ {[\Cref{prop:w1-hard}]}};
    \node[xp, font={\small\bfseries}, above=1.5cm of td, align=center] (vi) {vertex integrity\\ {[\Cref{thm:XP}]}};
    \node[nphard, right=3cm of cw] (tw) {treewidth \cite{Monien88}};
    \node[nphard, above right=1.1cm and 1.3cm of tw] (pw) {pathwidth \cite{Monien88}};

    \node[open, right=3cm of tw] (tcw) {tree cut width};
    \node[xp, above right=2cm and 1.5cm of tcw] (twd) {treewidth\\\(+\Delta\) \cite{Thilikos05_2}};
    \node[fpt, right=4.5cm of vc] (ml) {max leaf\\number};
    \node[fpt, below=1.5cm of ml] (bw) {bandwidth};
    \node[fpt, below=1.5cm of bw] (pwd) {pathwidth\\\(+ \Delta\)};
    \node[fpt, below=1.5cm of pwd] (cutw) {cutwidth\\\cite{Thilikos05_1}};

    \begin{scope}[every path/.style={->, semithick, shorten >=2pt, shorten <=2pt}]
    \path
      (vc) edge (rtc) edge (obtc)
      (rtc) edge (tc) edge (nd)
      (obtc) edge (tc) edge (cvdo)
      (ecc) edge (nd)
      (nd) edge (mw) edge (sd)
      (mw) edge (cw)
      (sd) edge (cw)
      (tw) edge (cw)
      (tcw) edge (tw)
      (twd) edge (tcw)
      (bw) edge (pwd)
      (ml) edge (bw)
      (cutw) edge (pw)
      (td) edge (pw) edge (sd)
      (cvd) edge (sd)
      (vi) edge (td)
      (cvdo) edge (cvd) edge (vi)
      (tc) edge (cvd) edge (mw)
      (pw) edge (tw)
      (pwd) edge (cutw)
      (cutw) edge (pwd);
    \draw[rounded corners] (bw) -| (twd);
    \end{scope}
  \end{tikzpicture}
  \caption{Different parameters and their relationship and results regarding parameterization of \cw.
    An arrow means that the pointed to parameter is more general and hardness propagates along arrows.
    Algorithmic results propagate in the reverse direction.
    All parameters marked with \protect\tikz[baseline=.05cm]\protect\draw[rounded corners=1pt, semithick, fill=T-Q-M2] rectangle +(.3, .3); admit an \FPT algorithm for \cw.
  \protect\tikz[baseline=.05cm]\protect\draw[rounded corners=1pt, semithick, fill=T-Q-M4!60] rectangle +(.3, .3); means there exists an \XP algorithm for \cw.
  \protect\tikz[baseline=.05cm]\protect\draw[rounded corners=1pt, semithick, fill=T-Q-M6] rectangle +(.3, .3); and  \protect\tikz[baseline=.05cm]\protect\draw[rounded corners=1pt, semithick, fill=T-Q-M9!60] rectangle +(.3, .3); mean that the corresponding parameterized problem is \(W[1]\)-hard and \NP-hard respectively. The remaining problems are open.
  The bold parameters are our contribution.
  This illustration is inspired by \cite[Figure 1]{gima23}.}
  \label{fig:cutwidth-fpt-net}
\end{figure}
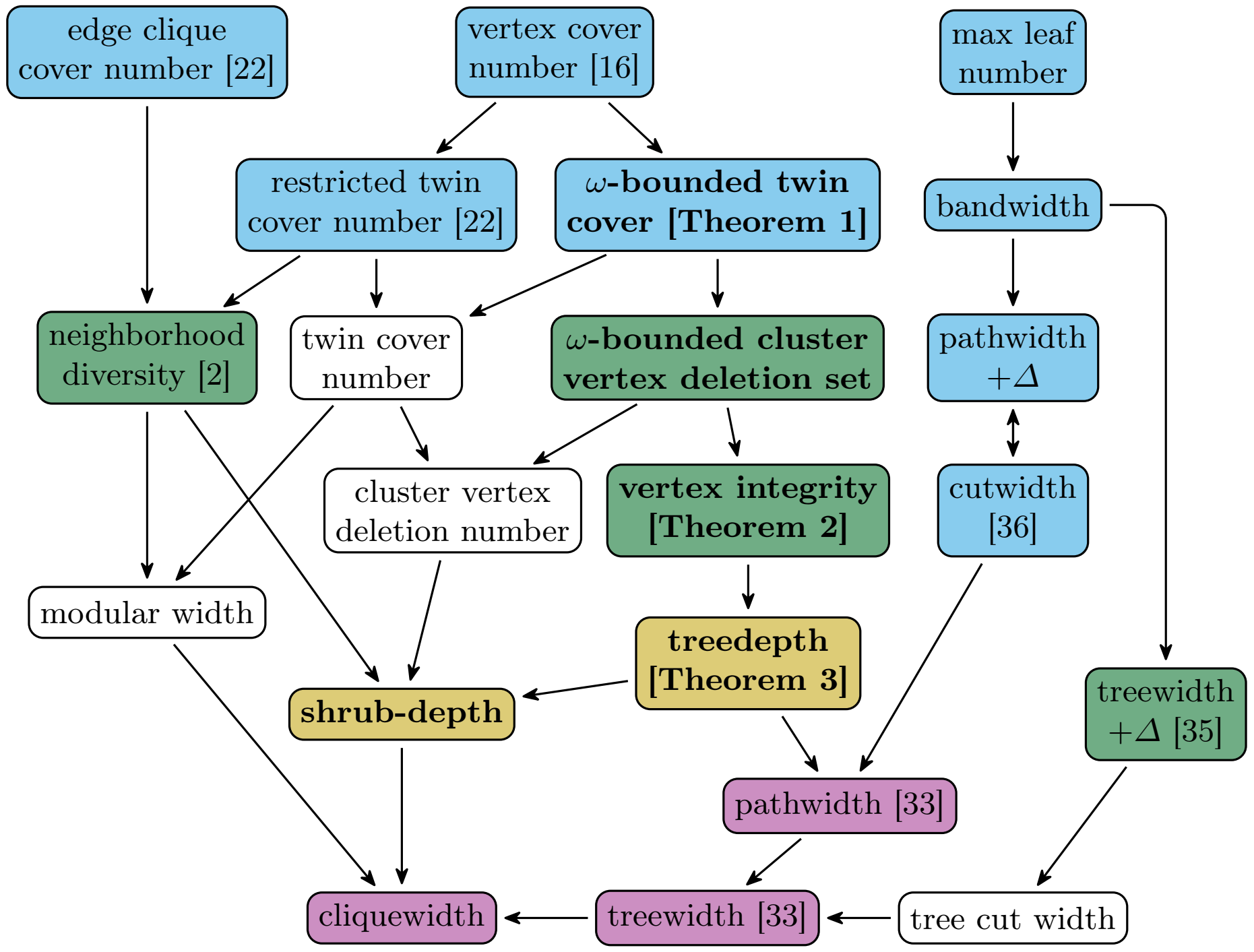

Due to the complexity of the problem, several results regarding approximation~\cite{Leighton88,Arora96,EvenNRS00,DiazPPS01,CaselDFKMS2019} and restricted graphs~\cite{Heggernes12,Yannakakis85,Heggernes11,Lengauer82,Nakano03,Harper64} exist. For an overview, we refer to~\cite{DiazPS02}.
Furthermore, parameterized approaches are of great interest.
In a fundamental work, Fellows, Lokshtanov, Misra, Rosamond, and Saurabh~\cite{FellowsLMRS08} prove that multiple layout problems including \cw are fixed parameter tractable (\FPT) when parameterized by the size of the minimum vertex cover.
Their $2^{2^k} n^{O(1)}$-time algorithm was improved by Cygan, Lokshtanov, Pilipczuk, Pilipczuk, and Saurabh~\cite{CyganLPPS14} to $2^k n^{O(1)}$ who also proved that this parameterization does not admit a polynomial kernel unless $\NP \subseteq \coNP/\mathrm{poly}$.
The problem is also \FPT when parameterized by the cutwidth value itself, which is equivalent to being \FPT when parameterized by pathwidth combined with the maximum degree~\cite{Thilikos05_1}. Furthermore, there exist \FPT-algorithms for the parameters edge clique cover number~\cite{gorzny22phd} and restricted twin cover~\cite{gorzny22phd}.
\cw is in \XP when parameterized by treewidth plus the maximum degree~\cite{Thilikos05_2} and neighborhood diversity~\cite{Bakken18}. The problem is \NP-hard when parameterized by treewidth or pathwidth, which also implies \NP-hardness when parameterized by cliquewidth~\cite{Monien88}.

The question whether \cw is \FPT parameterized by twin cover is of great interest and has gained some attention in the past.
While the removal of a vertex cover produces an independent set, the removal of a twin cover generates disjoint cliques where all vertices have the same neighborhood.
A formal definition for twin cover is given in \Cref{sec:prelims}.
Gorzny and Buss~\cite{GorznyB19} claim that \cw and \textsc{Imbalance} are both \FPT parameterized by twin cover, which was later declared as incorrect~\cite{gorzny21note}.
Their approach contracted the cliques of true twins into single vertices, which loses too much information~\cite{misra20}. Misra and Mittal then showed that \textsc{Imbalance} is in \XP when parameterized solely by twin cover and in \FPT when parameterized by \(\omega\)-bounded twin cover, i.e., where the maximum size of a clique is bounded by \(\omega\).
Gorzny~\cite{gorzny22phd} corrected and limited their result by showing that \cw is \FPT parameterized by \emph{restricted} twin cover, i.e., where the number of cliques with at least one edge is bounded.
Whether \cw can be solved in \FPT-time parameterized solely by twin cover remains open. We take a step into this direction by giving an \FPT algorithm for \cw parameterized by $\omega$-bounded twin cover. 
Notably, \(\omega\)-bounded twin cover and restricted twin cover both generalize vertex cover and are generalized by twin cover but are still incomparable. There exist families of graphs with bounded \(\omega\)-bounded twin cover number but unbounded restricted vertex cover and vice-versa.

In this work, we obtain new complexity results for five parameterizations (highlighted in bold in \Cref{fig:cutwidth-fpt-net}).
In \Cref{sec:twin cover}, we first strengthen a structural, not invalidated result from~\cite{GorznyB19} by proving that there exists a cutwidth-minimal ordering where the vertices of the cliques are placed contiguously and the cliques themselves follow a specific order. By fixing the twin cover and formulating an integer linear program (ILP), we obtain the following result.
\begin{restateable}{theorem}{thm:FPT}
    \cw is in \FPT when parameterized by $\omega$-bounded twin cover.
\end{restateable}
Furthermore, we explore more general parameters: In \Cref{sec:XP}, we propose an \XP algorithm for \cw parameterized by vertex integrity by designing a state-bounded dynamic programming algorithm over the connected components. Additionally, we prove $W[1]$-hardness for \cw parameterized by treedepth via a reduction from \textsc{MulticoloredClique} in \Cref{sec:treedepth}.
\begin{restateable}{theorem}{thm:XP}
    \cw is in \XP when parameterized by vertex integrity.
\end{restateable}

\begin{restateable}{theorem}{prop:w1-hard}
  \cw is \(W[1]\)-hard when parameterized by treedepth.
\end{restateable}
These results directly imply that $\cw$ is in \XP when parameterized by $\omega$-bounded cluster vertex deletion set and $W[1]$-hard when parameterized by shrub-depth.

\section{Preliminaries}\label{sec:prelims}
For $n\in \mathbb{Z}_{\ge1}$, we define $[n] \coloneqq \{1, 2, \ldots, n\}$ and $[n]_0 \coloneqq [n] \cup \{0\}$.
We denote an undirected graph without self-loops by $G=(V,E)$. 
For a subset of vertices $S \subseteq V$, we denote by $G[S]$ the subgraph induced by $S$, i.e., $G[S] \coloneqq (S, \{\{u,v\} \in E : u,v \in S\})$. Furthermore, we define $G-S \coloneqq G[V\setminus S], V(G) = V$ and $E(G) = E$. We use $\mathrm{cc}(G)$ to denote the set of maximal connected components of $G$.
For a subset of vertices $V' \subseteq V$, we denote the set of edges with exactly one end in $V'$ by \(\delta_G(V')\coloneqq\set{e\in E}[\abs{e\cap V'} = 1]\).

Given a linear ordering $\pi\in \Pi(G)$, we define \(V^\pi_i\coloneqq\set{v\in V}[\pi(v) \leq i]\) as the first \(i\) vertices in \(\pi\).
With this notation, we can use \(\delta_G\) to express the maximum cut of \(\pi\in\Pi(G)\) as \(\maxcut_G(\pi) = \max_{i\in[n-1]}\abs{\delta_G(V^\pi_i)}\).

We define the \emph{neighborhood} of a vertex $v \in V$ by $N(v) \coloneqq \{u \in V : \{u, v\} \in E\}$ and its \emph{closed neighborhood} by $N[v] \coloneqq N(v) \cup \{v\}$.
If all vertices in a subset $Y \subseteq V$ have the same closed neighborhood, i.e., for all $u,v \in Y$, it holds that $N[u] = N[v]$, we call the vertices of $Y$ a set of \emph{true twins}. Note that this property is an equivalence relation.

Many structural graph parameters are defined via a \emph{separator} $S \subseteq V$.
The removal of $S$ from $G$ leaves a subgraph $G - S$, with connected components $\mathrm{cc}(G-S)$.
Depending on the structural restrictions on $S$ and the components in $\mathrm{cc}(G-S)$, we obtain different parameters.

\begin{definition}[$\omega$-bounded twin cover]
    A separator $S \subseteq V$ is called a \emph{twin cover} if the remaining connected components $\mathrm{cc}(G-S)$ are sets of true twins, i.e., every $C \in \mathrm{cc}(G-S)$ induces a clique, and, for every pair $u,v \in V(C)$, we have $N(u) \cap S = N(v) \cap S$.
    Furthermore, $S$ is an $\omega$-bounded twin cover of size $k$ if it is a twin cover, $|S| \leq k$, and the number of vertices in each component does not exceed $\omega$, i.e., $|V(C)| \leq \omega$ for all $C \in \mathrm{cc}(G-S)$.
\end{definition}
For convenience, we usually write $\omega$-bounded twin cover while omitting the size $k$. However, we still refer to the definitions above.

A more general parameter only restricts the sizes of the separator and the components, without enforcing any internal structure on the components:
\begin{definition}[vertex integrity]
    The \emph{vertex integrity} $\mathrm{vi}(G)$ of a graph $G = (V,E)$ is the minimum sum of the size of a separator $S$ and the maximum size of a connected component in $G-S$, i.e., $\mathrm{vi}(G) = \min_{S\subseteq V} \abs{S} + \max_{C\in\mathrm{cc}(G-S)} \abs{V(C)}$.
\end{definition}

Finally, we define treedepth, which is even more general and is based on recursively removing vertices rather than analyzing connected components of a single separator.
\begin{definition}[treedepth]
  The treedepth $\mathrm{td}(G)$ of a graph \(G = (V,E)\) is
  \[
    \mathrm{td}(G) \coloneqq
    \begin{cases*}
      1 & if \(\abs{V} = 1\), \\
      1 + \min_{v\in V} \mathrm{td}(G - v) & if \(\abs{\mathrm{cc}(G)} = 1\) and \(\abs{V} > 1\), \\
      \max_{G'\in\mathrm{cc}(G)}\mathrm{td}(G') & otherwise.
    \end{cases*}
  \]
\end{definition}
We use (\faCut) to indicate that the proof of a lemma is provided in the appendix.

\section{An \FPT Algorithm for $\omega$-bounded Twin Cover}\label{sec:twin cover}

In this section, we present our \FPT algorithm, which conceptually extends the approach of Fellows et al.~\cite{FellowsLMRS08} for \cw parameterized by vertex cover. 
Their algorithm first guesses an ordering of the $k$ vertex cover vertices, reducing the problem to assigning the remaining independent vertices to the $k+1$ spaces (``locations'') between them.
Within each location, the vertices are optimally ordered. 
Grouping independent vertices by their neighborhoods bounds the number of distinct vertex types by $2^k$, enabling an ILP formulation with a number of variables bounded by a function of $k$.

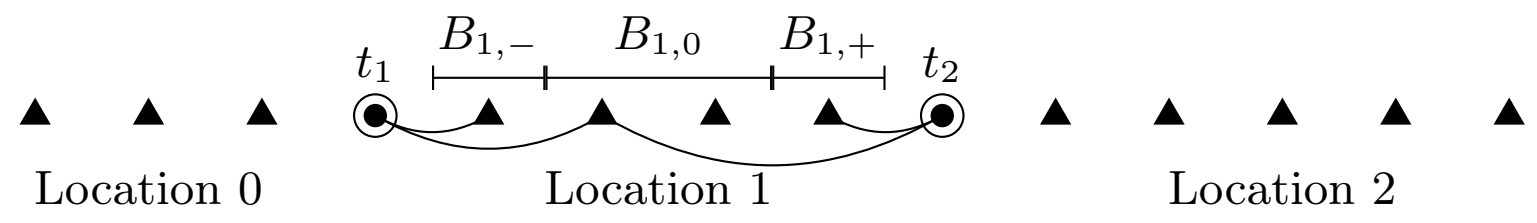
\begin{figure}[tb]
  \centering
  \begin{tikzpicture}
    \draw[fill] foreach \x in {0, 1, 2, 4, 5, 6, 7, 9, 10, ..., 13} { (\x*.75, 0) node[draw, fill, black, regular polygon, regular polygon sides=3, minimum size=6pt, inner sep=0pt] {} };

    \draw[fill] foreach \x [count=\i] in {3, 8} {(\x*.75, 0) circle (2pt) };
    \draw foreach \x [count=\i] in {3, 8} {(\x*.75, 0) circle (4pt) node [above=3pt] {\(t_{\i}\)}};
    \foreach \x [count=\i from 0] in {1, 5.5, 11} {
      \node[below=.25cm] at (\x*.75, 0) {Location \i};
    }
    \draw[|-|] (3.5*.75, .25) -- node[above] {\(B_{1, -}\)} ++(1*.75, 0);
    \draw[|-|] (4.5*.75, .25) -- node[above] {\(B_{1, 0}\)} ++(2*.75, 0);
    \draw[|-|] (6.5*.75, .25) -- node[above] {\(B_{1, +}\)} ++(1*.75, 0);
    \draw (3*.75, 0) to[bend right] (4*.75, 0);
    \draw (3*.75, 0) to[bend right] (5*.75, 0);

    \draw (5*.75, 0) to[bend right] (8*.75, 0);
    \draw (7*.75, 0) to[bend right] (8*.75, 0);
  \end{tikzpicture}
  \caption{Ordered set of vertices. Circled vertices are part of a twin cover; triangular vertices are cliques of true twins. For the cliques in location 1, the connections to vertices of the twin cover and the partition into \(B_{1,-}\), \(B_{1, 0}\), and \(B_{1, +}\) are shown.}
  \label{fig:location-example}
\end{figure}

Let $G=(V,E)$ be an undirected graph and $\{s_1,\dots,s_k\} = S \subseteq V$ a twin cover. Throughout this section, we assume that an ordering $\pi_S$ of the twin cover vertices is given. W.l.o.g., we say $\pi_S(s_1) = 1, \pi_S(s_2) = 2, \dots, \pi_S(s_k) = k$.
By a \emph{location}, we refer to a position for the remaining vertices $V \setminus S$ relative to $\pi_S$. Location 0 is on the left of vertex \(s_1\), location \(i\in [k-1]\) is between $s_i$ and $s_{i+1}$ and location $k$ is on the right of $s_k$ (see \Cref{fig:location-example}). 

We adapt the methodology from \cite{FellowsLMRS08} to $\omega$-bounded twin cover. More precisely, we aim to find an ordering for the vertices in every location.
However, considering every vertex individually is too expensive.
Using the following structural insight from Gorzny and Buss~\cite{GorznyB19}, we may treat the remaining cliques of true twins as units while still taking the sizes into account. 
\begin{theorem}[{Gorzny and Buss~\cite[Theorem 6]{GorznyB19}}]\label{prop:gorzny-structure}
    For any graph $G = (V,E)$, there exists a cutwidth-minimal ordering $\pi$ of $V$ such that each equivalence class of true twins appears consecutively in $\pi$.
\end{theorem}
For us, this theorem means that we can imagine a clique (\(C\in\mathrm{cc}(G-S)\)) as a single vertex and place these abstracted cliques into locations.
We note that when a clique of true twins is placed somewhere the order in which the vertices of this clique are placed does not matter.
Every order has the same cutwidth since all vertices have the same closed neighborhood.

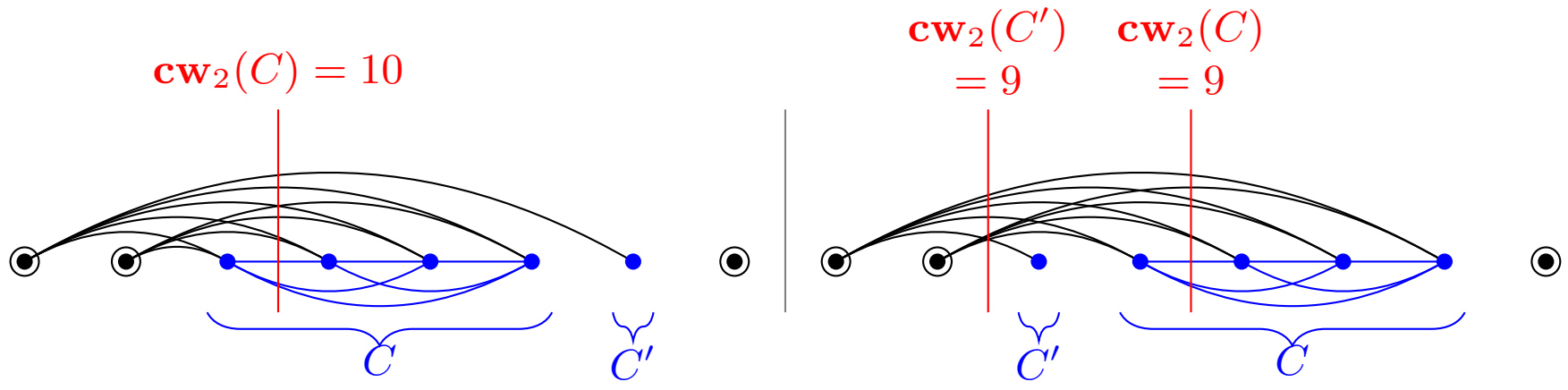
\begin{figure}[ht]
  \centering
  \begin{tikzpicture}[scale=.75]
  \begin{scope}
    \draw[fill] (0,0) circle (2pt);
    \draw[fill] (1,0) circle (2pt);
    \draw[fill] (7,0) circle (2pt);
    \draw foreach \x in {0,1,7} {(\x*1, 0) circle (4pt)};
    \draw [blue, decorate, decoration={brace, mirror, amplitude=7pt}] (1.8,-.5) -- (5.2,-.5) node[blue, midway, below=0.4em] {$C$};
    \draw [blue, decorate, decoration={brace, mirror, amplitude=6pt}] (5.8,-.5) -- (6.2,-.5) node[blue, midway, below=0.4em] {$C'$};
    \draw[blue] (2*1, 0) to (3*1, 0);
    \draw[blue] (2*1, 0) to[bend right] (4*1, 0);
    \draw[blue] (3*1, 0) to (4*1, 0);
    \draw[blue] (4*1, 0) to (5*1, 0);
    \draw[blue] (2*1, 0) to[bend right] (5*1, 0);
    \draw[blue] (3*1, 0) to[bend right] (5*1, 0);

    \draw (2*1, 0) to[bend right] (0*1, 0);
    \draw (2*1, 0) to[bend right] (1*1, 0);
    \draw (3*1, 0) to[bend right] (0*1, 0);
    \draw (3*1, 0) to[bend right] (1*1, 0);
    \draw (4*1, 0) to[bend right] (0*1, 0);
    \draw (4*1, 0) to[bend right] (1*1, 0);
    \draw (5*1, 0) to[bend right] (0*1, 0);
    \draw (5*1, 0) to[bend right] (1*1, 0);

    \draw (6*1, 0) to[bend right] (0*1, 0);
    
    \draw[fill, blue] (2,0) circle (2pt);
    \draw[fill, blue] (3,0) circle (2pt);
    \draw[fill, blue] (4,0) circle (2pt);
    \draw[fill, blue] (5,0) circle (2pt);
    \draw[fill, blue] (6,0) circle (2pt);

    \draw[red] (2.5, -0.5) -- +(0, 2)node[above] {\(\cwi[2]{C} = 10\)};
  \end{scope}
  \draw[gray] (7.5, -.5) -- +(0, 2);
  \begin{scope}[xshift=8cm]
    \draw[fill] (0,0) circle (2pt);
    \draw[fill] (1,0) circle (2pt);
    \draw[fill] (7,0) circle (2pt);
    \draw foreach \x in {0,1,7} {(\x*1, 0) circle (4pt)};
    \draw [blue, decorate, decoration={brace, mirror, amplitude=7pt}] (2.8,-.5) -- (6.2,-.5) node[blue, midway, below=0.4em] {$C$};
    \draw [blue, decorate, decoration={brace, mirror, amplitude=6pt}] (1.8,-.5) -- (2.2,-.5) node[blue, midway, below=0.4em] {$C'$};
    \draw[blue] (3*1, 0) to (4*1, 0);
    \draw[blue] (3*1, 0) to[bend right] (5*1, 0);
    \draw[blue] (4*1, 0) to (5*1, 0);
    \draw[blue] (5*1, 0) to (6*1, 0);
    \draw[blue] (3*1, 0) to[bend right] (6*1, 0);
    \draw[blue] (4*1, 0) to[bend right] (6*1, 0);

    \draw (3*1, 0) to[bend right] (0*1, 0);
    \draw (3*1, 0) to[bend right] (1*1, 0);
    \draw (4*1, 0) to[bend right] (0*1, 0);
    \draw (4*1, 0) to[bend right] (1*1, 0);
    \draw (5*1, 0) to[bend right] (0*1, 0);
    \draw (5*1, 0) to[bend right] (1*1, 0);
    \draw (6*1, 0) to[bend right] (0*1, 0);
    \draw (6*1, 0) to[bend right] (1*1, 0);

    \draw (2*1, 0) to[bend right] (0*1, 0);
    
    \draw[fill, blue] (2,0) circle (2pt);
    \draw[fill, blue] (3,0) circle (2pt);
    \draw[fill, blue] (4,0) circle (2pt);
    \draw[fill, blue] (5,0) circle (2pt);
    \draw[fill, blue] (6,0) circle (2pt);

    \draw[red] (3.5, -0.5) -- +(0,2)node[above, align=center] {\(\cwi[2]{C} \) \\ \( = 9\)};
    \draw[red] (1.5, -0.5) -- +(0,2)node[above, align=center] {\(\cwi[2]{C'}\) \\ \( = 9\)};
  \end{scope}
  \end{tikzpicture}
  \caption{The figure depicts the resulting ordering and the induced cutwidth $c$ for two cliques \(C, C'\in \mathrm{cc}(G-S)\) with \(\abs{V(C)} = 4, \abs{V(C')} = 1\) at location $i$ based on their rank compared to an optimal ordering based on \cref{prop:swap-is-good}.}
  \label{fig:counterexRank}
\end{figure}

We now establish a cutwidth-minimal ordering for cliques assigned to the same location.
In~\cite{FellowsLMRS08}, the vertices were ordered by their rank -- the difference between the number of neighboring vertices or twin cover vertices on the left and right. This does not suffice in our case (\Cref{fig:counterexRank}).
Simply reversing the order breaks for other examples, such as cliques of size one.
In fact, we have to consider the size of every clique while searching for a cutwidth-minimal ordering.
Formally, we want to find a \emph{location assignment} \(p : \mathrm{cc}(G-S) \to [k]_0\) that assigns every clique a location as well as an order for the cliques in every location.
To begin, notice that changing the order of two neighboring cliques in a location only changes the cuts in and between the cliques:
\begin{observation}\label{prop:swapping-is-local}
  Let \(\pi\) be an ordering of vertices where two cliques \(C_1, C_2\in\mathrm{cc}(G - S)\) appear consecutively.
  Let $s\in[n]$ be the first position of a vertex of \(C_1\) in \(\pi\) and \(e\in[n]\) be the  last position of a vertex of \(C_2\) in \(\pi\).
  Let \(\pi'\) be the ordering obtained by switching the positions of \(C_1\) and \(C_2\).
  Then,
  \begin{itemize}
  \item \(\delta_G(\set{v \in V}[\pi(v) < j]) = \delta_G(\set{v \in V}[\pi'(v) < j])\) for every \(j \leq s\) and
  \item \(\delta_G(\set{v \in V}[\pi(v) > j]) = \delta_G(\set{v \in V}[\pi'(v) > j])\) for every \(j\geq e\).
  \end{itemize}
\end{observation}
Thus, to evaluate a swap, we only need to consider cuts in and between the two cliques.

To determine whether a swap is beneficial, we calculate the maximum cut within a single clique. For this, we differentiate two kinds of edges: The edges that have at least one endpoint in the clique and the edges that have no endpoint in the clique.
Since the second kind of edges contribute equally to all cuts within the clique, we can ignore them when determining the clique's maximum internal cut. This \emph{maximum inner cut} depends on the location but is independent of other cliques. For a clique $C$ at location $i \in [k]_0$, it is given by
\begin{align*}
  \cwi{C} \coloneqq &\max_{x\in[\abs{V(C)}]_0} ((\abs{V(C)} - x)x + xR_{i}(C) + (\abs{V(C)} - x)L_{i}(C))\\
          = &\max_{x\in[\abs{V(C)}]_0} (- x^2 + (\abs{V(C)} + R_{i}(C) - L_{i}(C))x + \abs{V(C)}L_i(C))
\end{align*}
where \(L_i(C)\) (resp.\ \(R_i(C)\)) are the number of vertices from \(\set{s_1, \ldots, s_i}\) (resp.\ \(\set{s_{i + 1}, \ldots, s_k}\)) that are connected to the clique \(C\) (\Cref{fig:counterexRank}).
For more details see \cite[Section 3.7 and Lemma 3.7.4]{gorzny22phd}.

Next, we extend the analysis to two neighboring cliques \(C, C'\in\mathrm{cc}(G - S)\) in location \(i\).
If there are edges contributing to a cut in or between the two cliques and have no endpoint in either of the two cliques it has to contribute to every cut in or between.
Let \(E_o\) be the number of such edges.
Then, the maximum cut inside \(C\) and \(C'\) is given by
\[
  \cwi{C} + \abs{V(C')}L_i(C') + E_o \text{ and }
  \cwi{C'} + \abs{V(C)}R_i(C) + E_o
\]
respectively.
Swapping the order of \(C\) and \(C'\), the maximum cuts change to
\[
  \cwi{C'} + \abs{V(C)}L_i(C) + E_o \text{ and }
  \cwi{C} + \abs{V(C')}R_i(C') + E_o.
\]
This means, we can disregard \(E_o\), when evaluating whether a switch is beneficial.
To capture this idea, we write
\[
  \cwi{C, C'} \coloneqq \max(\cwi{C} + \abs{V(C')}L_{i}(C'), \cwi{C'} + \abs{V(C)}R_{i}(C))
\]
and introduce an order for the cliques based on this notion
\[
  C \lesssim_i C' \coloniff \cwi{C, C'} \leq \cwi{C', C}
\]
for \(C, C'\in\mathrm{cc}(G - S)\) and \(i\in[k]_0\).
The following lemma shows that swapping two neighboring cliques such that they abide \(\lesssim_i\) in location \(i\) does not increase the cutwidth of an ordering.
\begin{lemma}\label{prop:swap-is-good}
  Let \(\pi\) be an ordering of \(G\), and \(C, C'\in\mathrm{cc}(G - S)\) such that \(C\) is directly left to \(C'\) in location \(i\) and \(C' \lesssim_i C\).
  Then, swapping the order of \(C\) and \(C'\) does not increase the cutwidth of the ordering.
\end{lemma}
\begin{proof}
  By \Cref{prop:swapping-is-local} we know that cutwidth at every position before and after the cliques remains the same.
  Let \(s\) be first position of a vertex of \(C\) in \(\pi\) and \(e\) the last position of a vertex of \(C'\) in \(\pi\).
  Then, \(E_o \coloneqq \abs{\set{\set{v, u}\in E}[\pi(v) < s, e < \pi(u)]}\) is the number edges contributing to all cuts inside and between the cliques in \(\pi\) that have no endpoint in either clique.
  Since \(C' \lesssim_i C\), we know that
  \begin{multline*}
    \max(\cwi{C'} + \abs{V(C)}L_{i}(C), \cwi{C} + \abs{V(C')}R_{i}(C')) + E_o\\
    \leq
    \max(\cwi{C} + \abs{V(C')}L_{i}(C'), \cwi{C'} + \abs{V(C)}R_{i}(C)) + E_o.
  \end{multline*}
  Thus, changing the order does not increase the cutwidth. 
  \qed
\end{proof}
We would like to order the cliques in every location according to the predicate above.
Unfortunately, this predicate is not transitive due to problems with cliques where \(L_i(C) = R_i(C)\).
Let \(C\) be a clique with \(\abs{V(C)} = 1\) and \(L_i(C) = R_i(C) = 1\) and let \(C'\) be a clique with \(\abs{V(C')} = 1 \), \(L_i(C') = 1\), and \(R_i(C') = 0\). Notice that \(\cwi{C, C'} = 2 = \cwi{C', C}\).
The same holds for \(C'' \in \mathrm{cc}(G-S)\) with \(\abs{V(C'')} = 1\), \(L_i(C'') = 0\), and \(R_i(C'') = 1\).
However, we have \(\cwi{C', C''} < \cwi{C'', C'}\).

To circumvent this issue we first split each location in three parts \(B_{i, -}\), \(B_{i, 0}\), and \(B_{i, +}\) based on the rank (difference of \(L_i(C)\) and \(R_i(C)\)):
\[
  B_{i, -} \coloneqq \set{C\in\mathrm{cc}(G-S)}[p(C) = i, R_i(C) < L_i(C)] 
\]
for all \(i\in[k]_0\) and analogously, we define \(B_{i, 0}\) and \(B_{i, +}\) by replacing the \(<\) with \(=\) and \(>\), respectively (\Cref{fig:location-example}).
Furthermore, we give an order for each of the three parts.
The cliques in \(B_{i, -}\) and \(B_{i, +}\) may be placed in any order respecting \(\lesssim_i\).
The order of the cliques in \(B_{i, 0}\) does not matter.
In conclusion, we obtain the following lemma by iteratively applying \Cref{prop:swap-is-good}.

\begin{restateable}{lemma}{lem:structure}[\faCut]
  Let \(G = (V,E)\) be a graph, \(S\) a twin cover of size \(k\) of \(G\), and \(\pi^{*}\) cutwidth-minimal ordering of \(V\), where each equivalence class of true twins appears consecutively.
  Then any ordering \(\pi\) of \(V\) that
  \begin{itemize}
  \item agrees with \(\pi^{*}\) on \(S\) (\(\pi^*(s) \leq \pi^*(s') \iff \pi(s) \leq \pi(s')\) for all \(s, s'\in S\)),
  \item has the same location assignment \(p : \mathrm{cc}(G-S)\to [k]_0\),
  \item for every \(C\in\mathrm{cc}(G-S)\) the vertices of \(C\) appear consecutively in location \(p(C)\) of \(\pi\),
  \item for every \(i\in[k]_0\) the cliques are placed in the order \(B_{i, -}\), \(B_{i, 0}\), \(B_{i, +}\), and
  \item the cliques in \(B_{i, -}\), \(B_{i, +}\) respect the order \(\lesssim_i\)
  \end{itemize}
  is also optimal.
\end{restateable}

Now, we are ready to optimally determine the location assignment.
The general structure of the algorithm follows standard techniques for layout problems (e.g.~\cite{FellowsLMRS08,misra20}): we first compute and guess the ordering of an $\omega$-bounded twin cover. Using \Cref{lem:structure}, we can then formulate an ILP to optimally distribute the cliques to the locations. This ILP has $O(2^k \cdot \omega \cdot k)$ variables, thus by using the well-known algorithm by Lenstra~\cite{Lenstra83}, it can be solved in \FPT-time proving \Cref{thm:FPT}. The full details of the algorithm, the ILP formulation, and its correctness are deferred to \Cref{app:TCILP}.

\section{An \XP Algorithm for Vertex Integrity}\label{sec:XP}
Just by omitting the neighborhood restriction for the cliques, we loose the property (\Cref{prop:gorzny-structure}) that there always exists an optimal ordering where cliques are placed contiguously. For details see \Cref{sec:structure-breakdown}.
We therefore need a different approach.

Let a graph $G = (V,E)$ and an integer $k\in \ZZ_{\geq 0}$ be given. We can find an a separator $S$ with $|S|+\max_{C \in \mathrm{cc}(G-S)} |V(C)| \leq k$ in \FPT-time or decide that it does not exist~\cite{FellowsS89}.
In the following, we assume a separator $S \subseteq V$ is given such that $k$ is the vertex integrity of $G$.

To compute the optimal cutwidth of $G$, we propose a dynamic programming (DP) algorithm that constructs the linear ordering from left to right, placing one vertex at a time. At any given position, the algorithm must decide whether to place a vertex from the separator $S$ or a vertex from one of the connected components in $G-S$. Instead of tracking the exact permutation of all placed vertices, the DP only records the placed subset of $S$ and counts how many components of each structurally equivalent type currently have a specific subset of their vertices placed. As a key property, we use that only the already placed vertices matter for the current cut, not their exact order. Thus, multiple distinct sequences of vertex placements that result in the same set of placed vertices can be collapsed into a single DP state, where we only keep the minimum maximum-cut up to the current state. Like this, we iteratively build the layout until all vertices are placed.

We start with some definitions. Let \(C_1, C_2\in\mathrm{cc}(G - S)\) be connected components, and define the induced subgraphs \(G_1 \coloneqq G[V(C_1) \cup S]\) and \(G_2 \coloneqq G[V(C_2) \cup S]\).
\(C_1\) and \(C_2\) are defined as \emph{equivalent} if there exists a graph isomorphism (bijective mapping between graph vertices preserving adjacency) \(\varphi : V(G_1) \to V(G_2)\) between \(G_1\) and \(G_2\) such that \(\varphi(s) = s\) for all \(s\in S\).
We say that all equivalent components are of the same \emph{type} by choosing an arbitrary representative subgraph $T$ for the equivalence class.
Let $\cT$ be the set of all such component types.
For each type $T \in \cT$, let \(\cC_T\) be all connected components of \(G - S\) with type \(T\).
By the definition of type we have a graph isomorphism \(\varphi_C : V(T) \cup S \to V(C) \cup S\) between \(G[V(T) \cup S]\) and \(G[V(C) \cup S]\) for every \(T\in\cT\) and \(C\in\cC_T\).

We introduce an \emph{extended type} to manage the subset of vertices of a component that have already been placed:
\[
  \cT_\ext \coloneqq \set{(T, V_p)}[T\in\cT, V_p\subseteq V(T)].
\] 
Here, $V_p$ represents the subset of \emph{placed} vertices. 
Note that both, $|\cT|$ and $|\cT_\ext|$ are bounded by functions of $k$.
We provide more details for this in the proof of \Cref{prop:build-ordering-from-dp}. 

A \emph{configuration} is a function $f:\cT_\ext \to [n]_0$, such that 
    $\sum_{V_p \subseteq V(T)} f(T,V_p) = \abs{\cC_T}$
for all \(T\in\cT\), that tracks which vertices of \(G - S\) are placed grouped by type.
A \emph{state} describes all placed vertices up to position $i$ and is defined as a triple $(i,S_p,f)$, where $i \in [n]_0$ is a position in the layout, $S_p \subseteq S$ represents the separator vertices placed within the first $i$ positions, and $f$ is the current configuration, such that $i = |S_p| + \sum_{(T,V_p) \in \cT_\ext} f(T,V_p) \cdot |V_p|.$

A state \((i - 1, S'_p, f')\) is a predecessor of \((i, S_p, f)\) if we obtain \((i, S_p, f)\) from \((i - 1, S'_p, f')\) by placing one vertex at position \(i\).
More formally, when exactly one of the following holds:
\begin{itemize}
\item Separator vertex placed: There is a vertex \(s\in S_p\) such that \(S'_p = S_p\setminus\set{s}\) and \(f' = f\).
\item Connected component vertex placed: \(S'_p = S_p\) and there is an extended type \((T, V_p)\in\cT_\ext\) and a vertex \(v\in V_p\) such that
\[
  f'(T,V_p) = f(T,V_p) - 1
  \text{ and }
  f'(T, V_p\setminus\set{v}) = f(T, V_p\setminus\set{v}) + 1
\]
and \(f(T',V_p') = f'(T',V_p')\) for all other \((T', V_p')\in\cT_\ext\).
\end{itemize}
For a state $(i,S_p,f)$, we define the set of predecessor states $\Pred(i,S_p,f)$.

Finally, we adapt the definition of the size of a cut to this definition.
Recall that there are no edges between different connected components. Thus, adapting this to our definition of states, we obtain \(\delta : \mathcal P(S) \times [n]_0^{\cT_\ext} \to \bN\) with
\[
  \delta(S_p, f) \coloneqq \abs{\delta_{G[S]}(S_p)} + \smashoperator{\sum_{(T, V_p)\in\cT_\ext}}f(T, V_p)\cdot\abs[\big]{\delta_{G[V(T) \cup S]}(V_p \cup S_p)\setminus\delta_{G[S]}(S_p)}
\]
that calculates the cut based on the vertices placed in this state.
To show that this definition is sensible, we define the \emph{canonic states}.
Recall that for an ordering $\pi$ and \(i\in[n]_0\) the first \(i\) vertices of \(\pi\) are denoted by \(V^\pi_i\). 
We define \(S^\pi_i \coloneqq V^\pi_i \cap S\) and
\[
  f^\pi_i : \cT_{\ext} \to [n]_0,\;
  (T, V_p) \mapsto \abs[\big]{\set[\big]{C\in\cC_T : \varphi_C(V_p) = V^\pi_i \cap V(C)}}
\]
that describe the vertices placed up to position \(i\) by \(\pi\).
The following lemma shows that the sizes of the actual cuts equal the values from our abstraction.
\begin{restateable}{lemma}{prop:abstraction-equals-cut}[\faCut]
  For every ordering \(\pi\) and \(i\in[n]_0\) we have \(\abs{\delta_G(V^\pi_i)} = \delta(S_i^\pi, f_i^\pi).\)
\end{restateable}
To compute the optimal cutwidth, we define the DP table \(D[i, S_p, f]\), which stores the minimum maximum-cut in the first \(i\) positions over all orderings \(\pi\) such that \(S_p = S_i^\pi\) and \(f = f_i^\pi\) for every state \((i, S_, f)\).
We calculate the table entries as follows:
\begin{description}
\item[Base Case] At $i = 0$, the only state is the empty placement.
Thus, for $S_0 = \emptyset$ and $f_0$ with $f_0(T, \emptyset) = \abs{\cC_T}$ for all $T \in \cT$, and 0 otherwise, we set \(D[0, S_0, f_0] = 0\).

\item[Recursive Step]
For $i>0$, a separator subset $S_p\subseteq S$, and a configuration $f$, we minimize the maximum-cut over all possible predecessor states, and maximize that value against the current cut $\delta(S_p,f)$. Thus,
\begin{align*}
    D[i,S_p,f] = \max(\delta(S_P,f), \min_{(i-1,S_p',f')\in \Pred(i,S_p,f)} D[i-1,S_p',f']).
\end{align*}
\end{description}
The following lemmas show that the dynamic program calculates what we claim.
\begin{lemma}\label{prop:dp-correct}
  Let \(\pi\in\Pi(G)\) and \(i\in[n]_0\), we have \(D[i, S^\pi_i, f^\pi_i] \leq \max_{j\in[i]_0}\abs{\delta_G(V^\pi_j)}\).
\end{lemma}
\begin{proof}
  For \(i = 0\), we have \(D[i, S_i^\pi, f_i^\pi] = D[0, \emptyset, f_0] = 0 = \max_{j\in[0]_0}\abs{\delta_G(V^\pi_j)}\).

  For \(i > 0\), we have \((i - 1, S_{i - 1}^\pi, f_{i - 1}^\pi)\in\Pred(i, S_i^\pi, f_i^\pi)\) and thus by induction
  \begin{align*}
    D[i, S_i^\pi, f_i^\pi]
    &= \max(\delta(S_i^\pi,f_i^\pi), \min_{(i-1,S_p',f')\in \Pred(i,S_i^\pi,f_i^\pi)} D[i-1,S_p',f']) \\
    &\leq \max(\delta(S_i^\pi,f_i^\pi), D[i-1,S_{i - 1}^\pi,f_{i - 1}^\pi]) \\
    &\leq \max(\delta(S_i^\pi,f_i^\pi), \max_{j\in[i - 1]_0}(\abs{\delta_G(V^\pi_j)}) \\
    &=\max_{j\in[i]_0}(\abs{\delta_G(V^\pi_j)})
  \end{align*}
  where the last equality follows from \Cref{prop:abstraction-equals-cut}.
\qed
\end{proof}
Let $f_n$ be the configuration where all vertices have been placed, i.e., 
\begin{align*}
    f_n : \cT_\ext \to [n]_0, (T, V') \mapsto
  \begin{cases*}
    \abs{\cC_T} & if \(V' = V(T)\)\\
    0 & otherwise.
  \end{cases*}
\end{align*}
Since \(S = S_n^\pi\) and \(f_n = f_n^\pi\) for every ordering \(\pi\) we obtain the following corollary.
\begin{corollary}\label{prop:dp-is-cw}
  \(D[n, S, f_n] \leq \min_{\pi\in\Pi(G)}\maxcut_G(\pi) = \mathbf{cw}(G)\).
\end{corollary}
The next lemma shows that we can find a concrete ordering \(\pi\) such that \(\maxcut_G(\pi) \leq D[n, S, f_n]\) based on the dynamic program table in \XP time.

\begin{restateable}{lemma}{prop:build-ordering-from-dp}[\faCut]
  We can calculate the table \(D\) and find an ordering \(\pi\) such that \(\maxcut_G(\pi)\leq D[n, S, f_n]\) for every \(i\in[n]\) in \XP time.
\end{restateable}
The key insight to prove this lemma, is that when backtracking the states of our DP, we can place an arbitrary vertex that matches the transition of the DP (\Cref{prop:abstraction-equals-cut}).
Also, the number of states is bounded by \(n^{g(k)}\) for some computable function \(g\). 
With this, we have all necessary ingredients to prove that \cw is \XP parameterized by vertex integrity.

\restate{theorem}{thm:XP}
\begin{proof}
  We can find a separator that minimizes vertex integrity in \FPT time \cite{FellowsS89}.
  By \Cref{prop:build-ordering-from-dp}, we can calculate the dynamic program in \XP time.
  Furthermore, we obtain an ordering \(\pi\) from \Cref{prop:build-ordering-from-dp}.
  By \Cref{prop:dp-is-cw} we have \(\maxcut_G(\pi) \leq D[n, S, f_n] \leq \mathbf{cw}(G)\).
  \qed
\end{proof}

\section{\(W[1]\)-Hardness for Treedepth}\label{sec:treedepth}
To show that \cw is \(W[1]\)-hard parameterized by treedepth we adapt the reduction of Monien and Sudborough that shows that weighted \cw\footnote{Every edge has a weight and the maximum weighted cut in a linear order needs to be minimized.} is \NP-hard on trees with polynomial edge weights.
The key of their proof is a reduction from ILPs with special structure to weighted \cw on trees.
An NP-hard variant of \textsc{3-SAT} can be modeled as such an ILP which shows the NP-hardness of weighted \cw on trees with polynomial edge weights.
We will use a different problem, namely \textsc{MulticoloredClique}, to show the \(W[1]\)-hardness of \cw parameterized by treedepth.

We state the construction of Monien and Sudborough as the following Lemma which combines different results from their paper.
We only analyze the bounds, as they are not explicitly discussed in their paper.
\begin{restateable}{lemma}{prop:monien-reduction}[\faCut\;{\cite[Lemmas 2.3 and 2.4]{Monien88}}]
  Given an ILP with \(A\in\bZ_{\geq 0}^{m\times n}\) and \(b\in\bZ_{\geq 0}^m\) such that \(n\) is even and for every \(x\in\set{0, 1}^n\) with \(Ax = b\), we have \(\norm{x}_1 = \frac{n}{2}\).
  There exists an edge weighted tree \(T\) with \(O(nm)\) vertices, height \(O(m)\), and weights bounded by \(O(n^2m\Delta)\) where \(\Delta \coloneqq \max(\norm{A}_{\infty}, \norm{b}_{\infty})\).
  Furthermore, there is a value \(y\) such that \(\mathbf{cw}(T) \leq y\) if and only if there exists \(x\in\set{0, 1}^n\) with \(Ax = b\).
\end{restateable}
The bound on the weight is only a rough estimate but sufficient since we only need polynomially bounded weights.
To use this result for unweighted graphs, the weighted edges are substituted by parallel edges which in turn are replaced by 2-paths to obtain a graph without multiedges \cite[Lemma 2.1]{Monien88}.
The tree and the resulting graph have the same cutwidth since it is always optimal to move the midpoint of every created 2-path between the two end points.
Further, the treedepth of the graph is linear in height of the tree.

To show the \(W[1]\)-hardness of cutwidth parameterized by treedepth we reduce from \textsc{MulticoloredClique} instead of \textsc{3-SAT}. This problem is \(W[1]\)-hard when parameterized by the number of colors \cite{Fellows09}.
\begin{definition}[Multicolored Clique]
  Given a graph $G=(V,E)$ with a partition/coloring of the vertices \(V = V_1\sqcup \cdots \sqcup V_k\), the problem asks whether there exist a clique of size $k$ containing all colors, i.e., are there $v_1\in V_1,\dots,v_k \in V_k$ such that \(\set{v_i,v_j} \in E\) for all \(i,j \in [k]\) with $i \neq j$.
\end{definition}
To use this problem with the reduction from \cite{Monien88}, we model this problem as an ILP with \(O(k^2)\) constraints.
Then, the resulting tree has height and treedepth at most \(O(k^2)\).
The idea of the ILP formulation is that we have 0/1 variables for every vertex and every edge.
We select a vertex for every color and an edge between all pairs of colorclasses.
For every edge we make sure that the endpoints are selected vertices.
\begin{restateable}{lemma}{prop:multicolored-clique-to-ilp}[\faCut]
  Given an instance of \textsc{MulticoloredClique} \(G = (V = V_1 \sqcup \cdots \sqcup V_k, E)\).
  There is an ILP \(A\in\bZ_{\geq 0}^{m\times n}, b\in\bZ_{\geq 0}^n\) with \(n\in O(\abs{V} + \abs{E})\) variables, \(m\in O(k^2)\) constraints, and \(\max(\norm{A}_{\infty}, \norm{b}_{\infty})\in O(\abs{V} + \abs{E})\) such that there exists an \(x\in\set{0, 1}^n\) with \(Ax = b\) if and only if \(G\) is a yes-instance.
  Furthermore, \(n\) is even and for every \(x\in\set{0, 1}^n\) with \(Ax = b\), we have \(\norm{x}_1 = \frac{n}{2}\).
\end{restateable}

\restate{theorem}{prop:w1-hard}
\begin{proof}
  Let \(G = (V, E)\) be an instance of \textsc{MulticoloredClique} with \(k\) colors.
  Using \Cref{prop:monien-reduction} on the output of \Cref{prop:multicolored-clique-to-ilp}, we obtain an equivalent weighted \cw instance.
  The resulting graph is a tree with height \(O(k^2)\) and polynomial edge weights.
  Now, we replace the weighted edges by parallel 2-paths where the amount of paths is the weight of the replaced edge.
  This results in a equivalent \cw instance \cite{Monien88}.

  Next, we argue that the treedepth of this graph is bounded by \(O(k^2)\):
  Eliminating the vertices of the original tree layer by layer results in a graph which only contains the newly added midpoints of the 2-paths.
  All of these vertices are isolated.
  Thus, the number of the layers is at most the height of the tree plus one, i.e., the treedepth is \(O(k^2)\).

  Altogether, we have constructed a parameterized reduction from \textsc{MulticoloredClique} parameterized by the number of colors to \cw parameterized by the treedepth of the graph.
  The claim follows from the \(W[1]\)-hardness of \textsc{MulticoloredClique} \cite{Fellows09}.
  \qed
\end{proof}

\section{Conclusion}
While we gave an \FPT algorithm for \cw parameterized by $\omega$-bounded twin cover, the question of whether the problem is in \FPT parameterized solely by twin cover remains unresolved.
We refined the structural understanding of optimal solutions of \cw parameterized by twin cover.
Furthermore, we established that \cw is $W[1]$-hard when parameterized by treedepth and in \XP when parameterized by vertex integrity. For vertex integrity, it is currently unknown whether an \FPT algorithm exists. Since this parameter sits directly at the edge between our \XP and $W[1]$-hardness results, determining its exact complexity provides a interesting direction for future research. 
An overview of our results, as well as remaining open questions, can be seen in \Cref{fig:cutwidth-fpt-net}.

\begin{credits}
\subsubsection{\ackname} Funded by the Deutsche Forschungsgemeinschaft (DFG, German Research Foundation) – 528381760.
\end{credits}

\bibliography{references}

\clearpage
\appendix
\section{Omitted Proofs of \Cref{sec:twin cover}}\label{app:twin cover} 
\restate{lemma}{lem:structure}
\begin{proof}
  We start with the ordering given by \(\pi^{*}\).
  In this ordering the vertices of all equivalence classes appear consecutively.
  In the remaining proof we only swap neighboring equivalence classes in the same location and thus the constructed ordering will also have this property.
  
  Let \(C\in B_{i, -}\), \(C'\in B_{i, 0}\cup B_{i, +}\) be neighboring cliques at location $i$.
  We show that \(\cwi{C, C'} \leq \cwi{C', C}\) and thus if \(C'\) is left of \(C\) we can swap the cliques without increasing the cutwidth by \Cref{prop:swap-is-good}.

    We have $L_{i}(C)>R_{i}(C)$ and $L_{i}(C') \leq R_{i}(C')$ due to \(C\in B_{i, -}\) and \(C'\in B_{i, 0}\cup B_{i, +}\).
    Thus,
    \[
      \cwi{C} + \abs{V(C')}L_{i}(C') \leq \cwi{C} + \abs{V(C')}R_{i}(C')
    \]
    and
    \[
      \cwi{C'} + \abs{V(C)}R_{i}(C) < \cwi{C'} + \abs{V(C)}L_{i}(C)
    \]
    which implies \(\cwi{C, C'} \leq \cwi{C', C}\).

    In the same way, we can show a similar result for cliques from \(B_{i, 0}\) and \(B_{i, +}\).
    Therefore, we can swap neighboring cliques until we reach the desired partition without increasing the cutwidth. The obtained order satisfies the first four properties.
    Next, we show that \(\lesssim_i\) is transitive when restricted to cliques in \(B_{i, -}\).
    Let \(C, C'\in B_{i, -}\).
    Since \(\cwi{C} + \abs{V(C')}L_i(C') > \cwi{C} + \abs{V(C')}R_i(C')\) and \(\cwi{C'} + \abs{V(C)}R_i(C) < \cwi{C'} + \abs{V(C)}L_i(C)\), we have
    \begin{align*}
      C \lesssim_i C'
      &\iff \cwi{C} + \abs{V(C')}L_i(C') \leq \cwi{C'} + \abs{V(C)}L_i(C)\\
      &\iff\cwi{C} - \abs{V(C)}L_i(C) \leq \cwi{C'} - \abs{V(C')}L_i(C').
    \end{align*}
    Therefore, the relation \(C \lesssim_i C'\) is transitive when restricted to cliques in \(B_{i, -}\) and thus a total preorder (reflexive, transitive, and total) but notably not antisymmetric.
    The equivalence classes of the equivalence relation induced by \(\lesssim_i\) on \(B_{i, -}\) may contain multiple cliques.
    Equivalent cliques may be freely interchanged/swapped as \(C \lesssim_i C'\) and \(C' \lesssim_i C\) together imply \(\cwi{C, C'} = \cwi{C', C}\).

  This allows us to swap the cliques in \(B_{i, -}\) into any order respecting \(\lesssim_i\) while not increasing the cutwidth.
  Then, we can reach any order of cliques in \(B_{i, -}\) respecting \(\lesssim_i\) by swapping cliques in equivalence classes without increasing the cutwidth.

  Analogously, we can prove the same for \(\lesssim_i\) when restricted to cliques from \(B_{i, +}\) and we can order the cliques in \(B_{i, +}\) according to it.
  Cliques in \(B_{i, 0}\) can be swapped arbitrarily, since \(\cwi{C, C'} = \cwi{C', C}\) for all \(C, C'\in B_{i, 0}\). \qed
\end{proof}

\subsection{ILP Formulation and FPT Algorithm} \label{app:TCILP}
For the ILP, we guess the ordering of the vertices in a twin cover of size $k$ and then place the remaining vertices optimally in the locations before, between, or after the twin cover vertices. Using the strong structural result from the prior section, our algorithm achieves \FPT-time.  

In order to guess the ordering $\pi_S$ of the twin cover $S$, we first need to determine an $\omega$-bounded twin cover of size $k$. 
\begin{restateable}{lemma}{prop:getTwinCover}
    Given $\omega,k \in \ZZ_{\geq 0}$ and a graph $G = (V,E)$, we can, in \FPT-time, either find an $\omega$-bounded twin cover \(S\) of size at most $k$ of $G$, or decide that it does not exist.
\end{restateable}
\begin{proof}
    First, we compute all sets of true twins in time $O(|E|+|V|)$ using \cite[Lemma~5]{Ganian15}. Note that within a set, all vertices have the same closed neighborhood. Thus, all vertices of this set that are not part of the twin cover, form a single component. 
    So, for each set $C \in \mathrm{cc}(G)$ with more than $\omega$ vertices, i.e., $|V(C)|>\omega$, we add $|V(C)|-\omega$ vertices to the twin cover $S$. As all vertices have the same closed neighborhood, they can be chosen arbitrarily. This step runs in time $O(|V|^2)$. 
    
    Now, let $G'$ be the graph after removing all vertices in the current twin cover $S$. We now proceed as done by Ganian in \cite[Theorem 4]{Ganian15} for input parameter $k'=k-|S|$: In time $O(|E|+|V|)$, we remove all twin edges in $G'$ where a twin edge denotes an edge between two true twins. 
    Then in time $O(1.2738^{k'} + k'|V|)$, we compute the vertex cover of the remaining graph~\cite{ChenKX10}. By \cite[Lemma~2]{Ganian15}, this is a twin cover of $G'$. Therefore, we add these vertices to the twin cover $S$. If now $|S| > k$, there does not exist an $\omega$-bounded twin cover of size at most $k$ in $G$. Otherwise, return $S$.
    The total running time of the algorithm is $O(|E| + |V|^2 + 1.2738^{k'})$.
    \qed
\end{proof}
Given such an $\omega$-bounded twin cover $S$ with $|S| = k$, we guess the ordering $\pi_S$ of $S$ with an additional overhead of $k!$. In the following, we assume that such ordering is fixed and known.
By \Cref{lem:structure}, we can fix the order of the clique types within each location. We denote the ordering of the cliques at location $i$ by $\pi_i$. By fulfilling the stated properties, the lemma shows that the ordering is optimal w.r.t. a given location assignment.
The remaining task is to determine this location assignment optimally. To this end, we partition the cliques of $G-S$ into a set of types $\mathcal{T}$. We say a clique is of type $T = (\tau_N, \tau_s) \in \mathcal{T}$ if its neighborhood in the twin cover is exactly $\tau_N \subseteq S$ and it contains exactly $\tau_s \in [\omega]$ vertices. We formulate the assignment as an ILP to find how many cliques of each type occur in each location to minimize the maximum-cut.

Let the variable $c$ refer to the cutwidth value. Thus, our objective is 
\begin{align}
    \min c.
\end{align}
Additionally, we introduce the integer variables $x_{\type,i} \geq 0$ for all $\type \in \cT$ and $i\in [k]_0$, which denotes how many cliques of type $\type$ are placed at location $i$.
We also introduce a binary variable $y_{\type,i} \in \{0,1\}$ that equals $1$ if at least one clique of type $\type$ is placed at location $i$, and $0$ otherwise, i.e., we have 
\begin{align*}
    y_{\type,i} = \begin{cases}
        0, & \text{if } x_{\type,i} = 0 \\
        1, & \text{otherwise.}
    \end{cases}
\end{align*} 
Since $0 \leq x_{\type,i} \leq n$, this behavior is ensured by the following three constraints:
\begin{align}\label{eq:constraints_y}
    y_{\type,i} &\leq 1 & \text{for all } \type \in \cT, i \in [k]_0\nonumber\\
    x_{\type,i} &\leq n \cdot y_{\type,i} & \text{for all } \type \in \cT, i \in [k]_0\\
    y_{\type,i} &\leq x_{\type,i} & \text{for all } \type \in \cT, i \in [k]_0\nonumber
\end{align}

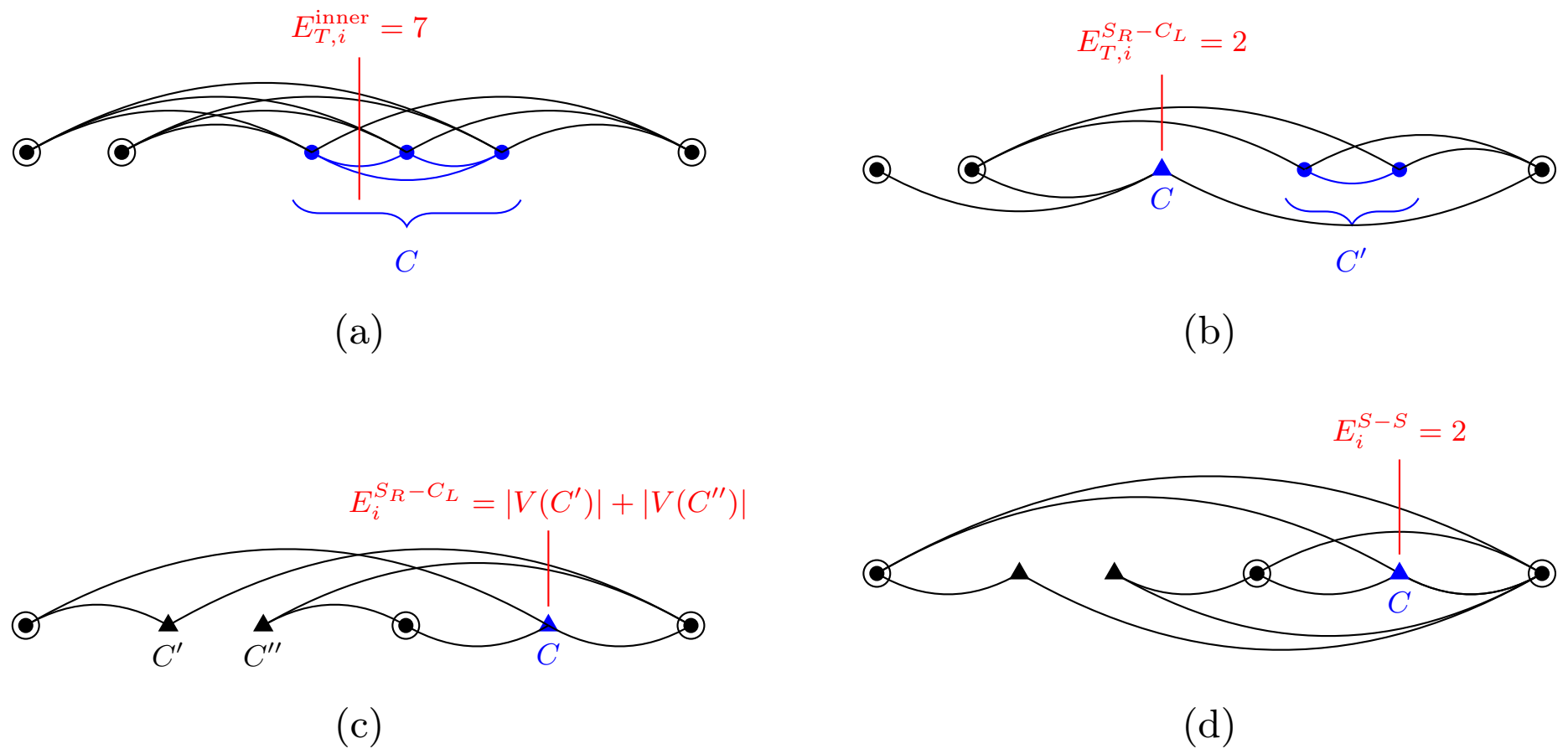
\begin{figure}[t]
\centering

\subcaptionbox{
\label{fig:a}}[0.45\textwidth]{
\centering
\begin{tikzpicture}[scale=0.75, transform shape]
        \draw[fill] (0,0) circle (2pt);
        \draw[fill] (1,0) circle (2pt);
        \draw[fill, blue] (3,0) circle (2pt);
        \draw[fill, blue] (4,0) circle (2pt);
        \draw[fill, blue] (5,0) circle (2pt);
        \draw[fill] (7,0) circle (2pt);
        \draw foreach \x in {0,1, 7} {(\x*1, 0) circle (4pt)};
        \draw [blue, decorate, decoration={brace, mirror, amplitude=6pt}] (2.8,-0.5) -- (5.2,-0.5) node[blue, midway, yshift=-2em] {$C$};
        \draw[blue] (3*1, 0) to[bend right] (4*1, 0);
        \draw[blue] (3*1, 0) to[bend right] (5*1, 0);
        \draw[blue] (4*1, 0) to[bend right] (5*1, 0);
    
        \draw (3*1, 0) to[bend right] (0*1, 0);
        \draw (3*1, 0) to[bend right] (1*1, 0);
        \draw (7*1, 0) to[bend right] (3*1, 0);
        \draw (4*1, 0) to[bend right] (0*1, 0);
        \draw (4*1, 0) to[bend right] (1*1, 0);
        \draw (7*1, 0) to[bend right] (4*1, 0);
        \draw (5*1, 0) to[bend right] (0*1, 0);
        \draw (5*1, 0) to[bend right] (1*1, 0);
        \draw (7*1, 0) to[bend right] (5*1, 0);

        \draw[red] (3.5, -0.5) -- (3.5,1)node[above] {\({E_{\type,i}^\text{inner}} = 7\)} ;
      \end{tikzpicture}

}
\hfill
\subcaptionbox{
\label{fig:b}}[0.45\textwidth]{
\centering
\begin{tikzpicture}[scale=0.75, transform shape]
        \draw[fill] (0,0) circle (2pt);
        \draw[fill] (1,0) circle (2pt);
        \draw[fill, blue] (4.5,0) circle (2pt);
        \draw[fill, blue] (5.5,0) circle (2pt); 
        \draw[fill] (7,0) circle (2pt);
        \draw foreach \x in {0,1, 7} {(\x*1, 0) circle (4pt)};
        
        \draw [blue, decorate, decoration={brace, mirror, amplitude=6pt}] (4.3,-0.3) -- (5.7,-0.3) node[blue, midway, yshift=-2em] {$C'$};
        
        \draw (3*1, 0) to[bend left] (0*1, 0);
        \draw (3*1, 0) to[bend left] (1*1, 0);
        \draw (3*1, 0) to[bend right] (7*1, 0);
        \draw (4.5*1, 0) to[bend left] (7*1, 0);
        \draw (5.5*1, 0) to[bend left] (7*1, 0);
        \draw[blue] (4.5*1, 0) to[bend right] (5.5*1, 0);
        \draw (4.5*1, 0) to[bend right] (1*1, 0);
        \draw (5.5*1, 0) to[bend right] (1*1, 0);

        \node[
            draw,
            fill,
            blue,
            regular polygon,
            regular polygon sides=3,
            minimum size=6pt,
            inner sep=0pt
          ] at (3,0) {};
        \node[blue] at (3,-0.3) {$C$};
        
        \draw[red] (3, 0.2) -- (3,1.0)node[above] {\(E_{\type,i}^{S_R-C_L} = 2\)} ;

      \end{tikzpicture}

}

\vspace{1em}

\subcaptionbox{
\label{fig:c}}[0.45\textwidth]{
\centering
\begin{tikzpicture}[scale=0.75, transform shape]
        \draw[fill] (0,0) circle (2pt);
        \draw[fill] (4,0) circle (2pt);
        \draw[fill] (7,0) circle (2pt);
        \node[
            draw,
            fill,
            black,
            regular polygon,
            regular polygon sides=3,
            minimum size=6pt,
            inner sep=0pt
        ] at (1.5,0) {};
        \node[] at (1.5,-0.3) {$C'$};
        
        \node[
            draw,
            fill,
            black,
            regular polygon,
            regular polygon sides=3,
            minimum size=6pt,
            inner sep=0pt
          ] at (2.5,0) {};
        \node[] at (2.5,-0.3) {$C''$};

        \node[
            draw,
            fill,
            blue,
            regular polygon,
            regular polygon sides=3,
            minimum size=6pt,
            inner sep=0pt
          ] at (5.5,0) {};
        \node[blue] at (5.5,-0.3) {$C$};
          
        \draw foreach \x in {0,4, 7} {(\x*1, 0) circle (4pt)};

        \draw[black] (0*1, 0) to[bend left] (1.5*1, 0);
        \draw[black] (2.5*1, 0) to[bend left] (4*1, 0);
        \draw[black] (1.5*1, 0) to[bend left] (7*1, 0);
        \draw[black] (2.5*1, 0) to[bend left] (7*1, 0);

        \draw[black] (5.5*1, 0) to[bend right] (0*1, 0);
        \draw[black] (5.5*1, 0) to[bend left] (4*1, 0);
        \draw[black] (7*1, 0) to[bend left] (5.5*1, 0);

        \draw[red] (5.5, 0.2) -- (5.5,1)node[above] {\({E_i^{S_R-C_L}} = |V(C')| + |V(C'')|\)} ;

      \end{tikzpicture}
}
\hfill
\subcaptionbox{
\label{fig:d}}[0.45\textwidth]{
\centering
\begin{tikzpicture}[scale=0.75, transform shape]
        \draw[fill] (0,0) circle (2pt);
        \draw[fill] (4,0) circle (2pt);

        \draw[fill] (7,0) circle (2pt);
        \draw foreach \x in {0,4, 7} {(\x*1, 0) circle (4pt)};
        \draw (7*1, 0) to[bend right] (0*1, 0);
        \draw (7*1, 0) to[bend right] (4*1, 0);

        \draw (0*1, 0) to[bend right] (1.5*1, 0);
        \draw (2.5*1, 0) to[bend right] (4*1, 0);
        \draw (5.5*1, 0) to[bend right] (7*1, 0);
        \draw (7*1, 0) to[bend left] (1.5*1, 0);
        \draw (7*1, 0) to[bend left] (2.5*1, 0);
        \draw (5.5*1, 0) to[bend right] (7*1, 0);
        \draw (5.5*1, 0) to[bend right] (0*1, 0);
        \draw (5.5*1, 0) to[bend left] (4*1, 0);        

        \node[
            draw,
            fill,
            black,
            regular polygon,
            regular polygon sides=3,
            minimum size=6pt,
            inner sep=0pt
          ] at (1.5,0) {};
          \node[
            draw,
            fill,
            black,
            regular polygon,
            regular polygon sides=3,
            minimum size=6pt,
            inner sep=0pt
          ] at (2.5,0) {};
          \node[
            draw,
            fill,
            blue,
            regular polygon,
            regular polygon sides=3,
            minimum size=6pt,
            inner sep=0pt
          ] at (5.5,0) {};        
          \node[blue] at (5.5,-0.3) {$C$};

          \draw[red] (5.5, 0.2) -- (5.5,1.2)node[above] {\(E_i^{S-S} = 2\)} ;

      \end{tikzpicture}
}

\caption{The figure illustrates the different edge types that need to be considered when calculating the maximum-cut within clique $C$ at location $i$. \Cref{fig:a} illustrates $E_{\type,i}^{\text{inner}}$ for $C \in B_{i,-}$. \Cref{fig:b} illustrates clique $C\in B_{i,-}$ of type $\type$ and $C'\in B_{i,0}$ of type $\type'$ at location $i$. In \cref{fig:c}, edges between $S$ and cliques at locations $\neq i$ are considered. \Cref{fig:d} depicts edges between vertices of the twin cover. Note that $E_i^{S-C_\type}$ is not depicted but the visualization is similar to \Cref{fig:b}. In general, vertices of the twin cover $S$ are denoted as \symTC, one clique as \symClique \ and individual vertices of a clique as \symCliqueVert. Accordingly, vertices and cliques at the considered location are colored \textcolor{blue}{blue}. Note that an edge between cliques (\symClique) and vertices are multiple edges. 
}
\label{fig:graphs}
\end{figure}

Now, we determine the maximum-cut that occurs when cutting through a clique of a given type $\type = (\tau_N, \tau_s) \in \cT$ at a specific location $i \in [k]_0$. Analogously to $L_i(C)$ and $R_i(C)$, we define $L_i(\type)$ and $R_i(\type)$ as the number of twin cover vertices to the left and right, respectively, that are connected to a clique of type $\type$ at location $i$. Furthermore, we assume that the maximum inner cut $\cwi{C}$ is precomputed for all cliques and locations.

To ensure we account for all relevant edges exactly once, recall that the graph $G-S$ is a disjoint union of cliques. Therefore, any edge in $G$ belongs to exactly one of three categories: (1) both endpoints are in $S$, (2) one endpoint is in $S$ and the other in $G-S$, or (3) both endpoints are strictly within the same clique in $G-S$.

\begin{itemize}
    \item \textbf{Inner Cut ($E_{\type,i}^{\text{inner}}$):} The here considered edges have at least one vertex within this clique (\Cref{fig:a}). By definition, $\cwi{C}$ captures all such edges. If no such clique is assigned to this location ($y_{\type,i}=0$), this value is 0. Hence,
    \begin{align*}
        E_{\type,i}^{\text{inner}} \coloneqq y_{\type,i} \cdot \cwi{C}.
    \end{align*}
    \item \textbf{Twin Cover -- Other cliques of type $\type$ at location $i$ ($E_{\type,i}^{S-C_\type}$):} These are edges connecting $S$ to the remaining $x_{\type,i} - y_{\type,i}$ cliques of the identical type at $i$. Because they share the same rank and neighborhood, the maximum cut occurs when $C$ is placed at the outermost boundary of this type group (leftmost if $L_i(\type) > R_i(\type)$, or rightmost otherwise):
    \begin{align*}
        E_{\type,i}^{S-C_\type} \coloneqq (x_{\type,i}-y_{\type,i}) \cdot \tau_s \cdot \max(L_i(\type), R_i(\type)). 
    \end{align*}
    \item \textbf{Twin Cover -- Cliques of different type $\type' \neq \type$ at location $i$ ($E_{\type,i}^{S - C_{T'}}$):} These are edges crossing $C$ that connect $S$ to cliques of type $T'$ at location $i$ (\Cref{fig:b}). Based on $\pi_i$ derived from \Cref{lem:structure}, these are edges from $S$ left of $i$ to cliques placed right of $\type$, and from $S$ right of $i$ to cliques placed left of $\type$. Let $E_{\type,i}^{S_L - C_R} \coloneqq \sum_{\type' : \pi_i(\type) < \pi_i(\type')} \tau_s' \cdot x_{\type',i} \cdot L_i(\type')$ and $E_{\type,i}^{S_R - C_L} \coloneqq \sum_{\type' : \pi_i(\type') < \pi_i(\type)} \tau_s' \cdot x_{\type',i} \cdot R_i(\type')$. Then,\begin{align*}
        E_{\type,i}^{S - C_{T'}} \coloneqq E_{\type,i}^{S_L - C_R} + E_{\type,i}^{S_R - C_L}.
    \end{align*}
    \item \textbf{Twin Cover -- Cliques at other locations ($E_i^{S - C_{\text{ext}}}$):} These are edges crossing location $i$ that connect $S$ to cliques at any other location $i' \neq i$ (\Cref{fig:c}). Thus,
    \begin{align*}
        E_i^{S - C_{\text{ext}}} \coloneqq \sum_{\type' \in \cT} \left( \sum_{i'=i+1}^k \tau_s' \cdot x_{\type', i'} \cdot L_i(\type') + \sum_{i'=0}^{i-1} \tau_s' \cdot x_{\type', i'} \cdot R_i(\type') \right).
    \end{align*}
    \item \textbf{Twin Cover -- Twin Cover ($E_i^{S-S}$):} These are edges connecting a twin cover vertex on the left of $i$ to one on the right of $i$ (\Cref{fig:d}). This value is constant for fixed $\pi_S$:
    \begin{align*}
        E_i^{S-S} \coloneqq |\{\{s,s'\} \in E \colon s,s' \in S, \pi_S(s) \leq i < \pi_S(s')\}|.
    \end{align*}
\end{itemize}

Since the vertices of the cliques are placed consecutively, edges solely between vertices of cliques $C' \in \mathrm{cc}(G-S)$ with $C' \neq C$ do not cross clique $C$. Same applies to all edges between vertices of cliques at location $j \in [k]_{<i}$ (respectively $j \in [k]_{>i}$) to vertices of the twin cover to the left (respectively to the right) of $C$. Additionally by definition of the twin cover, we do not have any edges connecting distinct cliques.
Thus, the linear constraints bounding the cutwidth $c$ is defined by the components stated above:
\begin{align}\label{eq:constraint_c}
    c &\geq E_{\type,i}^{\text{inner}} + E_{\type,i}^{S-C_\type} + E_{\type,i}^{S - C_{\type'}} + E_i^{S - C_{\text{ext}}} + E_i^{S-S} &  \text{for all } \type \in \cT, i \in [k]_0.
\end{align}

Finally, we need to ensure that we assign the cliques that actually appear in our input graph $G$, i.e., that the sum of all cliques of a type $\type \in \cT$ over all $k+1$ locations equals the number of their appearances in the input graph $G$. The following constraint ensures this property, where $W_\type$ denotes the number of occurrences of a clique of type $\type$ in $G$:
\begin{align}
    \sum_{i=0}^{k}{x_{\type,i}} &= W_\type & \text{for all } \type \in \cT.\label{eq:local constr}
\end{align}

In total this results in an ILP with $4(2^k \cdot \omega \cdot (k+1)) + 2^k\cdot \omega = O(2^k \cdot \omega \cdot k)$ constraints and $2(2^k \cdot \omega \cdot (k+1))+1 = O(2^k \cdot \omega \cdot k)$ variables. 
The largest coefficient within the constraints is $\max(n,2^k \cdot \omega \cdot (k+1))$ and the variables can be upper bounded by $n$.
\begin{lemma}\label{lem:ILPCorrectness}
    Let $G=(V,E)$ be a graph and $S=\{s_1,\dots,s_k\}$ an $\omega$-bounded twin cover. There exists a twin cover ordering $\pi_S$ such that the minimum value of $c$ satisfying the ILP constraints \eqref{eq:constraints_y}, \eqref{eq:constraint_c}, and \eqref{eq:local constr} with respect to $\pi_S$ equals the minimal cutwidth of $G$. 
\end{lemma}
\begin{proof}
    Let $\mathbf{cw}(G)$ denote the optimal cutwidth of $G$. Let $\pi^*$ be an optimal ordering of $G$ such that it fulfills the structural properties of \Cref{lem:structure} and $\maxcut_G(\pi^*) = \mathbf{cw}(G)$. Such an ordering maps directly to a feasible ILP assignment. Additionally, we can derive an optimal ordering from an optimal ILP solution.
    
    Let $\pi_S$ be the ordering of the twin cover vertices induced by $\pi^*$. We will show that for this $\pi_S$, the minimum value $c^*$ satisfying the ILP constraints \eqref{eq:constraints_y}, \eqref{eq:constraint_c}, and \eqref{eq:local constr} equals $\mathbf{cw}(G)$. We prove this by showing $c^* \leq \mathbf{cw}(G)$ and $c^* \geq \mathbf{cw}(G)$ separately.

    \begin{enumerate}
        \item[$\leq:$] By \Cref{lem:structure}, we can assume w.l.o.g. that $\pi^*$ places the vertices of each clique contiguously and orders the cliques as follows:
        \begin{itemize}
            \item cliques belonging to $B_{i,-}$ are ordered according to $\lesssim_i$ and placed leftmost in location $i$,
            \item cliques belonging to $B_{i,+}$ are ordered according to $\lesssim_i$ and placed rightmost in location $i$, and
            \item cliques belonging to $B_{i,0}$ are placed in between $B_{i,-}$ and $B_{i,+}$ while the order does not matter.
        \end{itemize}

        For each $\type \in \cT$ and $i \in [k]_0$, let $x_{\type,i}^*$ be the number of cliques of type $\type$ assigned to location $i$ by $\pi^*$. Set $y_{\type,i}^* = 1$ if $x_{\type,i}^*\geq 1$, and $0$ otherwise. By Lemma 10, every clique is assigned to exactly one location in the ordering $\pi^*$. This ensures that $\sum_{i=0}^k x_{\type,i}^* = W_\type$, satisfying \eqref{eq:local constr}. By definition of $y_{\type,i}^*$, \eqref{eq:constraints_y} also holds.

        By definition of the edge classes, the sum $E_{\type,i}^{\text{inner}} + E_{\type,i}^{S-C_\type} + E_{\type,i}^{S - C_{\type'}} + E_i^{S - C_{\text{ext}}} + E_i^{S-S}$ is the maximum-cut of a clique of type $\type \in \cT$ at location $i$ in $\pi^*$. 
        Since $\maxcut_G(\pi^*)$ is the global maximum over all positions in the ordering, we have
        \begin{align*}
            \maxcut_G(\pi^*) \geq E_{\type,i}^{\text{inner}} + E_{\type,i}^{S-C_\type} + E_{\type,i}^{S - C_{\type'}} + E_i^{S - C_{\text{ext}}} + E_i^{S-S}
        \end{align*}
        for all $\type\in \cT$ and $i \in [k]_0$.
        Note that the maximum-cut of a clique can be within or at the boundary of this clique. For all empty locations, the maximum-cut at that location is $E_i^{S - C_{\text{ext}}} + E_i^{S-S}$, as there are no cliques at that location and the other summands equal 0. Thus, we still obtain 
        \begin{align*}
            \maxcut_G(\pi^*) &\geq E_i^{S - C_{\text{ext}}} + E_i^{S-S} \\
            &= E_{\type,i}^{\text{inner}} + E_{\type,i}^{S-C_\type} + E_{\type,i}^{S - C_{\type'}} + E_i^{S - C_{\text{ext}}} + E_i^{S-S}.
        \end{align*}
        Thus, substituting $c = \maxcut_G(\pi^*)$ yields a solution to \eqref{eq:constraint_c}. 
        Since $c^*$ is the minimum feasible value for $c$, it follows that $c^* \leq c = \maxcut_G(\pi^*) = \mathbf{cw}(G)$.

        \item[$\geq:$] 
        Let $(x^*,y^*,c^*)$ be an optimal solution to \eqref{eq:constraints_y}, \eqref{eq:constraint_c}, and \eqref{eq:local constr} with respect to $\pi_S$. We construct an ordering $\pi$ of $G$ as follows:
        First, place all vertices of $S$ according to $\pi_S$. Then, for each $i \in [k]_0$ and $\type \in \cT$, place exactly $x_{\type,i}^*$ cliques of type $\type$ at location $i$. Order the cliques as described in \Cref{lem:structure}. Since the cliques consist of true twins and using \Cref{prop:gorzny-structure}, the vertices of a clique can now be placed contiguously in arbitrary order.

        Since the variables satisfy \eqref{eq:local constr}, the sum of assigned cliques of type $\type \in \cT$ over all locations is $W_T$. Thus, the assignment places every vertex of $G-S$ exactly once. Since also the vertices in $S$ are placed exactly once, this forms a complete linear ordering of $V$.

        By \Cref{lem:structure}, the global maximum-cut of $\pi$ must occur either within or directly adjacent to a clique $C$ assigned to some location $i$, or strictly between two twin cover vertices when the location does not have any assigned cliques.
        In the first case, the maximum-cut is defined by the right-hand side of \eqref{eq:constraint_c}. Since $c^*$ is an upper bound of this sum for all $\type\in \cT$ and $i \in [k]_0$, we have $\maxcut_G(\pi) \leq c^*$.
        In the second case, the cut equals $E_i^{S - C_{\text{ext}}} + E_i^{S-S}$. Observe that for any type $T \in \mathcal{T}$, the right-hand side of \eqref{eq:constraint_c} includes these terms alongside additional edge counts ($E_{\type,i}^{\text{inner}}$, $E_{\type,i}^{S-C_\type}$, and $E_{\type,i}^{S - C_{\type'}}$). Because these additional terms are always non-negative, the constraint evaluated for any arbitrary type $T$ ensures $c^* \geq E_i^{S - C_{\text{ext}}} + E_i^{S-S}$. Thus, we again have $\maxcut_G(\pi) \leq c^*$.

        Since $\pi$ is a valid ordering of $G$, its maximum-cut cannot be strictly less than the optimal cutwidth $\mathbf{cw}(G)$. Thus, $\mathbf{cw}(G) \leq \maxcut_G(\pi) \leq c^*$.
    \end{enumerate}
    Combining both bounds yields $c^* = \mathbf{cw}(G)$, completing the proof.
    \qed
\end{proof}

To solve the ILP, we rely on the well-known fixed-parameter tractability of ILPs parameterized by the number of variables $p$. While Lenstra~\cite{Lenstra83} and Kannan~\cite{Kannan87} originally established this for ILP feasibility, their framework can be adapted via a binary search over the objective value to solve the optimization version. Specifically, an ILP can be solved optimally in time $p^{O(p)}\cdot (|I| + \log(M_x))\log(M_xM_a)$, where $I$ denotes the input instance, $|I|$ is its encoding length and $M_x$ and $M_a$ are upper bounds on the absolute values of the variables and coefficients, respectively ~\cite{CyganFKLMPPS15}.

Since our formulation has $p = O(2^k \cdot \omega \cdot k)$ variables, we can find the optimal solution in time $(2^k \cdot \omega \cdot k)^{O(2^k \cdot \omega \cdot k)} \cdot \text{poly}(|I|)$. Altogether, we obtain an \FPT-algorithm for \cw parameterized by $\omega$-bounded twin cover.

\restate{theorem}{thm:FPT}
\begin{proof}
    By \Cref{prop:getTwinCover}, we can find an $\omega$-bounded twin cover of size $k$ in time $O(m+n^2+ 1.2738^k)$. Having this twin cover $S$, we try all $k!$ orderings $\pi_S$ of its vertices and solve the corresponding ILP in time $(2^k \cdot \omega \cdot k)^{O(2^k \cdot \omega \cdot k)} \cdot \text{poly}(|I|)$.
    In total, this results in a running time of
    \begin{align*}
        O(m+n^2+ 1.2738^k) + k! \cdot (2^k \cdot \omega \cdot k)^{O(2^k \cdot \omega \cdot k)} \cdot \text{poly}(|I|) \\
        \qquad = O(2^k \omega k)^{O(2^k \omega k)} \cdot \text{poly}(|I|)
    \end{align*}
    which is \FPT parameterized by $k$ and $\omega$.
    By \Cref{lem:ILPCorrectness}, there exists a twin cover ordering $\pi_S$ for which the ILP computes the minimum cutwidth. By exhaustively testing all possible orderings $\pi_S$, we thus find the optimal cutwidth.
    \qed
\end{proof}

\section{Omitted Proofs of \Cref{sec:XP}}
\subsection{Why the Approach for \(\omega\)-bounded Twin Cover is Difficult to Generalize}\label{sec:structure-breakdown}
\begin{figure}[t]
\centering
\begin{tikzpicture}
    \draw (0, 0) circle (3.5pt)
      -- ++(1, 0)
      -- ++(1, 0)
      -- ++(1, 0) circle (3.5pt)
      -- ++(-.5, -0.866)
      -- ++(-1, 0) circle (3.5pt)
      -- ++(.5, 0.866) ++(-.5, -0.866)
      -- ++(-.5, 0.866) ++(.5, -0.866)
      -- ++(-1, 0)
      -- ++(-.5, 0.866);
    \fill (0, 0) circle (2pt)
      ++(1, 0) circle (2pt)
      ++(1, 0) circle (2pt)
      ++(1, 0) circle (2pt)
      ++(-.5, -0.866) circle (2pt)
      ++(-1, 0) circle (2pt)
      ++(-1, 0) circle (2pt);

    \draw[gray] (4, -1.25) -- +(0, 1.5);
    
    \draw (5, -0.433) circle (3.5pt)
      -- ++(1, 0)
        ++(1, 0)
      -- ++(1, 0) circle (3.5pt)
      -- ++(1, 0)
        ++(1, 0)
      -- ++(1, 0) circle (3.5pt);

    \fill (5, -0.433) circle (2pt)
      ++(1, 0) circle (2pt)
      ++(1, 0) circle (2pt)
      ++(1, 0) circle (2pt)
      ++(1, 0) circle (2pt)
      ++(1, 0) circle (2pt)
      ++(1, 0) circle (2pt);

    \draw (5, -0.433) to[bend left] +(2, 0);
    \draw (6, -0.433) to[bend left] +(2, 0);
    \draw (7, -0.433) to[bend left] +(2, 0);
    \draw (8, -0.433) to[bend left] +(2, 0);
    \draw (9, -0.433) to[bend left] +(2, 0);
  \end{tikzpicture}
\caption{
A graph with a minimal \(\omega\)-bounded cluster vertex deletion set of size 3 with \(\omega = 2\).
The 2-bounded cluster vertex deletion set is marked by circles.
Removing these three vertices leaves one 2-clique and two 3-cliques. 
On the right a cutwidth-minimal ordering of the graph achieving a cutwidth of three.
Every ordering that places the 2-clique cosecutively has cutwidth at least four.}
\label{fig:clustervertexdeletion}
\end{figure}
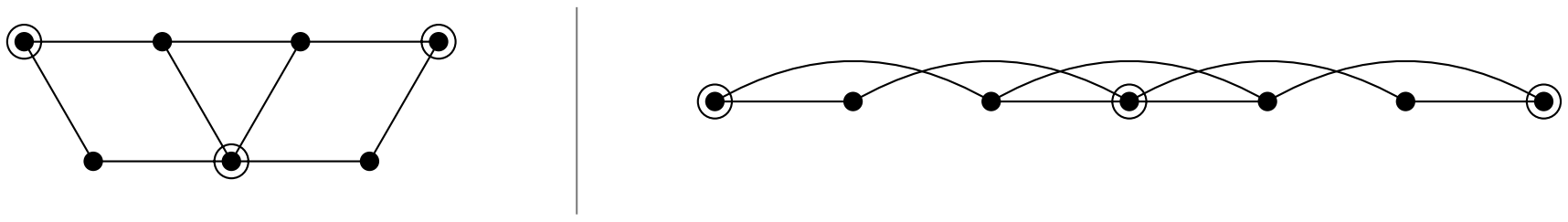

One might intuitively expect that a similar property holds for the cliques obtained by the removal of a cluster vertex deletion set.
This would allow us to again treat cliques as single units, assign them to locations, and order them via a result similar to \Cref{lem:structure}.
However, this is not the case.
An example graph that demonstrates this property is given in \Cref{fig:clustervertexdeletion}.

A similar approach would still be possible if we can show that there is an optimal solution in which the vertices of connected components appear consecutively in every location.
We were not able to show such a structural result or find a counter example yet.

\subsection{Omitted Proofs}
\restate{lemma}{prop:abstraction-equals-cut}
\begin{proof}
  We distinguish two kinds of edges:  Edges between the vertices of \(S\) and edges that are incident to at least one vertex in \(G - S\).
  The second kind of edge can be considered on a per connected component basis.
  Since \(V^\pi_i \cap S = S_i^\pi\), we have
  \begin{align*}
    \abs[\big]{\delta_G(V^\pi_i)}
    &= \abs[\big]{\delta_{G[S]}(V^\pi_i \cap S)} + \abs[\big]{\delta_{(V, \set{e\in{E} : \abs{e\cap S}\leq 1})}(V^\pi_i)}\\
    &=\abs{\delta_{G[S]}(S_i^\pi)} + \smashoperator{\sum_{C\in\mathrm{cc}(G - S)}}\,\abs[\big]{\delta_{G[V(C) \cup S]}\big(V^\pi_i \cap (V(C) \cup S)\big)\setminus\delta_{G[S]}(S_i^\pi)}.
  \end{align*}
  \(G[S]\) is a subgraph of \(G[V(C) \cup S]\) for every \(C\in\mathrm{cc}(G - S)\).
  To avoid counting the cut edges in \(G[S]\) multiple times we count them separately and remove them from every \(\delta_{G[V(C)\cup S]}\).
  We further analyze the second kind of edges:
  \begin{align*}
    \MoveEqLeft\smashoperator{\sum_{C\in\mathrm{cc}(G - S)}}\,\abs[\big]{\delta_{G[V(C) \cup S]}\big(V^\pi_i\cap (V(C) \cup S)\big)\setminus\delta_{G[S]}(S_i^\pi)}\\
    &=\sum_{T\in\cT}\sum_{V_p\subseteq V(T)}\abs[\big]{\set[\big]{C\in\cC_T : \varphi_C(V_p) = V^\pi_i \cap V(C)}}\\
    &\qquad\cdot\abs[\big]{\delta_{G[V(T) \cup S]}(V_p \cup S_i^\pi)\setminus\delta_{G[S]}(S_i^\pi)}\\
    &=\abs{\delta_{G[S]}(S_i^\pi)} + \smashoperator{\sum_{(T, V_p)\in\cT_\ext}}f_i^\pi(T, V_p) \cdot\abs[\big]{\delta_{G[V(T) \cup S]}(V_p \cup S_i^\pi)\setminus\delta_{G[S]}(S_i^\pi)}.
  \end{align*}
  Combining both shows that
  \begin{align*}
    \abs{\delta_G(V^\pi_i)}
    &=\abs{\delta_{G[S]}(S_i^\pi)} + \smashoperator{\sum_{(T, V_p)\in\cT_\ext}}f_i^\pi(T, V_p) \cdot\abs[\big]{\delta_{G[V(T) \cup S]}(V_p \cup S_i^\pi)\setminus\delta_{G[S]}(S_i^\pi)}\\
    &= \delta(S_i^\pi, f_i^\pi). \tag*{\qed}
  \end{align*}
\end{proof}

\restate{lemma}{prop:build-ordering-from-dp}
\begin{proof}
  We first argue the time needed to calculate the dynamic program.
  To that end, we start by bounding the number of states.
  Since $|S| \leq k$ and every component $C\in\mathrm{cc}(G - S)$ satisfies $|V(C)| \leq k$, the number of internal edges (edges connecting a vertex in $C$ with another vertex in $C$) and external neighborhood combinations (edges connecting a vertex in $C$ with a vertex in $S$) is bounded by a function of $k$.
  Specifically, there are at most $2^{{k}\choose{2}}$ possibilities for the internal edges of $C$ and at most $2^{k^2}$ possibilities for the edges between $C$ and $S$.
  Hence,
  \[
    |\cT| \leq k\cdot 2^{{{k}\choose{2}} + k^2} = 2^{O(k^2)}.
  \]
  Since $|V(T)| \leq k$, the number of extended types is bounded by $\cT_\ext \leq 2^k|\cT| \leq 2^{O(k^2)}$.
  When bounding the number of states, we have that $i \in [n]_0$ and that the number of subsets of $S$ is at most $2^k$.
  With $\cT_\ext \leq 2^k|\cT|$, we obtain that the total number of configurations $f$ is bounded by $(n+1)^{|\cT_\ext|}$.
  Thus, the dynamic table has at most $n\cdot 2^k \cdot n^{2^{O(k^2)}}$ entries.
  Since we can calculate the recursive step in the dynamic program in \XP time, we can calculate the entire table in \XP time.  

  Next, we discuss how to obtain the ordering \(\pi\).
  By the definition of recurrence of the dynamic program, there are states
  \[
      (0, S_0,f_0),(1, S_1,f_1),\dots,(n, S_n,f_n)
  \]
  such that \((i - 1, S_{i - 1},f_{i - 1})\in\Pred(i, S_i, f_i)\) for all \(i\in[n]\) and \(D[i, S_i, f_i]\leq D[n, S, f_n]\) for all \(i\in [n]_0\).
  To find these states, we backtrack through the dynamic programming table, which can be done in \XP time.
  We construct \(\pi\) by placing vertices from 1 to \(n\).
  For \(i\in[n]_0\), let \(V_i\) be the vertices we placed up to and including position \(i\) (\(V_0 = \emptyset\)).
  Let \(i\in[n]\).
  We consider the difference between \((i - 1, S_{i - 1},f_{i - 1})\) and \((i, S_{i},f_{i})\).
  By the definition of \(\Pred\), we have to consider two cases:
  \begin{itemize}
  \item Separator vertex placed: There is a vertex \(s\in S_i\) such that \(S_{i - 1} = S_i\setminus\set{s}\) and \(f_{i - 1} = f_i\).
  We set \(\pi(s) = i\) and \(V_i \coloneqq V_{i - 1} \cup \set{s}\).
  \item Connected component vertex placed: \(S_{i - 1} = S_{i}\) and there is an extended type \((T, V_p)\in\cT_\ext\) and a vertex \(v\in V_p\) such that
  \[
    f_i(T,V_p) =  f_{i - 1}(T,V_p) + 1
    \text{ and }
    f_i(T, V_p\setminus\set{v}) = f_{i - 1}(T, V_p\setminus\set{v}) - 1.
  \]
  There is a connected component \(C\in\cC_T\) such that \(\varphi_C(V_p\setminus\set{v}) = V(C) \cap V_{i - 1}\).
  Such a connected component always exists since by our construction we ensure that \(\abs{\set{C'\in\cC_T}[\varphi_{C'}(V_p \setminus \set{v}) = V(C') \cap V_{i - 1}]} = f_{i - 1}(T, V_p\setminus\set{v}) = f_i(T, V_p\setminus\set{v}) + 1 \geq 1\).
  We set \(\pi(\varphi_C(v)) = i\) and \(V_i \coloneqq V_{i - 1} \cup \set{\varphi_C(v)}\).
  \end{itemize}
  Since we keep track of the placed vertices in \(V_i\), no vertex is placed twice. Thus, \(\pi\) is a well-defined ordering.
  Furthermore, \(S_i^\pi = S_i\) and \(f_i^\pi = f_i\) for all \(i\in[n]_0\).
  Lastly, we show that \(\maxcut_G(\pi) \leq D[n, S, f_n]\).
  We have \(\delta(S_i, f_i) \leq D[i, S_i, f_i] \leq D[n, S, f_n]\) for every \(i\in[n]_0\).
  Due to \Cref{prop:abstraction-equals-cut}, we can bound the maximum cut in \(\pi\):
  \begin{align*}
    \maxcut_G(\pi)
    &= \max_{i\in[n - 1]}\abs{\delta_G(V^\pi_i)} \\
    &= \max_{i\in[n - 1]}\delta(S_i^\pi, f_i^\pi) \\
    &= \max_{i\in[n - 1]}\delta(S_i, f_i)\\
    &\leq D[n, S, f_n]. \tag*{\qed}
  \end{align*}
\end{proof}

\section{Omitted Proofs of \Cref{sec:treedepth}}
\restate{lemma}{prop:monien-reduction}
\begin{proof}
  In the proof of Lemma 2.3 the ILP is transformed so that it fulfills the preconditions of Lemma 2.4. We do not explicitly state these four steps, nor the proof. However, we analyze how the transformations in Lemma 2.3 change the properties of the matrix and how this results in a weighted tree in Lemma 2.4 of \cite{Monien88}.
  The remaining parts are proven in the original paper.
  \begin{description}
  \item[Lemma 2.3] Let \(\Delta_i\) be a bound on the largest absolute entries in the transformed matrix and   right-hand side after the $i$-th transformation and let $n_i$ and $m_i$ be the corresponding dimensions    of the transformed matrix.
    \begin{enumerate}
    \item Since the sum of rows are added to the values and two new columns are added, we have \(\Delta_1 \leq 2n\Delta\), \(n_1 = n + 2\), and \(m_1 = m\).
    \item Two columns are added with entries bounded by \(2\Delta_1\) and the right-hand side is adjusted accordingly.
    This yields \(\Delta_2 \leq 3\Delta_1 \leq 6n\Delta\), \(n_2 = n_1 + 2 = n + 4\), and \(m_2 = m_1 = m\).
    \item Large values are added to entries and the right-hand side to order the rows.
    The resulting ILP has \(\Delta_3 \leq \Delta_2 + \frac{1}{2}n_2m_2(\Delta_2 + 1) \leq 6n\Delta + \frac{1}{2}(n + 4)m(6n\Delta + 1)\in O(n^2m\Delta)\), \(n_3 = n_2 = n + 4\), and \(m_3 = m_2 = m\).
    \item The entries in the first row are doubled.
    We have \(\Delta_4 \leq 2\Delta_3\in O(n^2m\Delta)\), \(n_4 = n_3 = n + 4\), and \(m_4 = m_3 = m\).
    \end{enumerate}
  \item[Lemma 2.4] The constructed tree is a collection of \(O(n_4) \leq O(n)\) paths of length \(O(m_4) \leq O(m)\) joined together at a common root.
    The edge weights are at most \(2\Delta_4 \leq O(n^2m\Delta)\).
  \end{description}
  Combining the lemmas yields the desired result.\qed
\end{proof}

\restate{lemma}{prop:multicolored-clique-to-ilp}
\begin{proof}
  Let $E_{\set{i,j}}$ be the set of edges with one endpoint in $V_i$ and one in $V_j$.
  For all vertices $v \in V$ we introduce a variable $x_v$ and for all edges $e \in E$, we introduce a variable $x_e$.
  Further, we assume that the vertices are numbered from $1$ to $n$ and we obtain the number via a bijection $\mathrm{num} : V \to [\abs{V}]$.
  We formulate the following ILP with \(A\in\bZ_{\geq 0}^{m\times n}\) and \(b\in\bZ_{\geq 0}^n\)
  \begin{align*}
      \sum_{v \in V_i} x_v &= 1 & \text{for all } i \in [k]\\
      \sum_{e \in E_{\set{i,j}}} x_e &= 1 & \text{for all } i,j \in [k] \text{ with } i < j\\
      \sum_{v \in V_i} \mathrm{num}(v) \cdot x_v + \smashoperator{\sum_{\substack{e \in E_{\set{i, j}}\\v\in e \cap V_i}}}\; (\abs{V} - \mathrm{num}(v)) \cdot x_e &= \abs{V} & \text{for all } i,j \in [k]\text{ with } i \ne j\\
    x_v, x_e&\in\set{0, 1} & \text{for all } v\in V, e\in E
  \end{align*}
  where \(n = \abs{V} + \abs{E}\), \(m = O(k^2)\), and \(\max(\norm{A}_{\infty}, \norm{b}_{\infty})\leq \abs{V}\).
  The first constraint ensures that for every color exactly one vertex is selected and the second ensures the same for the edges between color classes.
  The third constraint guarantees that the endpoints of every selected edge are selected.

  When \(G\) has a multicolored clique of size \(k\) then this clique contains exactly one vertex from every color class.
  Further, between clique vertices of different color classes there is an edge.
  We set the corresponding variables in the ILP to one and all other variables to zero.
  The first two constraints are satisfied because we select exactly one vertex/edge for every class/between two classes.
  The third constraint is fulfilled since all selected edges are between vertices of the clique.

  If we have a solution for the ILP, we have selected exactly one vertex for every color class and exactly one edge between different color classes due to first two constraints.
  Thus, in the third constraint both sums have exactly one variable set to one.
  Rearranging the constraints shows
  \[
    \sum_{v \in V_i} \mathrm{num}(v) \cdot x_v = \sum_{\substack{e \in E_{\set{i, j}}\\v\in e \cap V_i}} \mathrm{num}(v) \cdot x_e
  \]
  for all \(i,j \in [k]\) with \(i \ne j\).
  Since \(\mathrm{num}\) is a bijection, the selected edges are between the selected vertices.
  The ILP solution selects edges between every pair of color classes and together with the third constraint this implies that the selected vertices form a multicolored clique of size \(k\).

  This ILP may not fulfill the condition that \(n\) is even and every solution \(x\in\set{0, 1}^n\) with \(Ax = b\) has \(\norm{x}_1 = \frac{n}{2}\).
  We modify the ILP such that the conditions are fulfilled.
  To this end, we define a new matrix and right-hand side
  \[
    A' \coloneqq
    \begin{pmatrix}
      A & 0\\
      1 \cdots 1 & 1 \cdots 1
    \end{pmatrix}\in\bZ_{\geq 0}^{m\times 2n}\quad\text{and}\quad
    b' \coloneqq
    \begin{pmatrix}
      b\\
      n
    \end{pmatrix}\in\bZ_{\geq 0}^m.
  \]
  This system is equivalent to the previous system since the newly added columns do not interact with the old system and are always satisfiable.
  Furthermore, this system has an even number \(2n\) of variables and every solution has exactly \(n = \abs{V} + \abs{E}\) variables set to one. \qed
\end{proof}
\end{document}